\documentclass[10pt]{article}
\usepackage{fancyhdr}
\usepackage{extramarks}
\usepackage{amsmath}
\usepackage{amsthm}
\usepackage{amsfonts}
\usepackage{siunitx}
\usepackage{tikz}
\usepackage[plain]{algorithm}
\usepackage{algpseudocode}
\usepackage{multirow}
\usepackage{booktabs}
\usepackage{graphicx}
\usepackage{subfigure}
\usepackage[colorlinks,linkcolor=black,anchorcolor=black,citecolor=black,urlcolor=blue]{hyperref}
\usepackage{amsmath,bm}
\usepackage{booktabs}
\usepackage{mathtools}
\usepackage{amssymb}
\usepackage{caption}
\usepackage{capt-of}
\usepackage{mciteplus}
\usepackage{cite}
\usepackage{mathrsfs}
\usepackage[title,titletoc,toc]{appendix}
\usepackage{xr}
\usepackage{parskip}
\usepackage{soul}
\usepackage{textcomp}
\usepackage[colaction]{multicol}
\usepackage[switch]{lineno}
\usepackage{lipsum}
\usepackage{etoolbox}
\usepackage{longtable}
\usepackage{array}
\usepackage{tablefootnote}
\usepackage{multirow}
\newcolumntype{C}[1]{>{\centering\arraybackslash}p{#1}}
\usetikzlibrary{automata,positioning}
\usetikzlibrary{automata,positioning}

\begin{document}

\title{Prediction and mitigation of mutation threats to COVID-19 vaccines and antibody therapies}
 
\author{Jiahui Chen$^1$,  Kaifu Gao$^{1,\dagger}$, Rui Wang$^{1,\dagger}$, 
 and Guo-Wei Wei$^{1,2,3}$\footnote{
Corresponding author.		E-mail: weig@msu.edu} \\
$^1$ Department of Mathematics, \\
Michigan State University, MI 48824, USA.\\
$^2$ Department of Electrical and Computer Engineering,\\
Michigan State University, MI 48824, USA. \\
$^3$ Department of Biochemistry and Molecular Biology,\\
Michigan State University, MI 48824, USA. \\
$^\dagger$First three authors contributed equally. 
}
\date{\today} 

\maketitle


\begin{abstract}
 Antibody therapeutics and vaccines are among our last resort to end the raging COVID-19 pandemic. They, however, are prone to over 5,000 mutations on the spike (S) protein uncovered by a \href{https://users.math.msu.edu/users/weig/SARS-CoV-2_Mutation_Tracker.html}{Mutation Tracker} {based on over 200,000 genome isolates}. It is imperative to understand how mutations would impact vaccines and antibodies in the development. In this work, we first study the mechanism, frequency, and ratio of mutations on the S protein  that is   the common target of most COVID-19 vaccines and antibody therapies. Additionally, we build a library of { 56} antibody structures and analyze their 2D and 3D characteristics. Moreover, we predict the mutation-induced binding free energy (BFE) changes for the complexes of S protein and antibodies or ACE2. By integrating genetics, biophysics, deep learning, and algebraic topology, we reveal that most of 462 mutations on the receptor-binding  domain (RBD) will weaken the binding of S protein and antibodies and disrupt the efficacy and reliability of antibody therapies and vaccines. A list of 31 vaccine escape mutants is identified, while many other disruptive mutations are detailed as well.  
We also unveil that about 65\% existing RBD mutations, including those variants recently found in the United Kingdom (UK)  and South Africa,  will strengthen the binding between the S protein and human angiotensin-converting enzyme 2 (ACE2), resulting in more infectious COVID-19 variants. We discover the disparity between the extreme values of RBD mutation-induced BFE strengthening and weakening of the bindings with antibodies and angiotensin-converting enzyme 2 (ACE2), suggesting that SARS-CoV-2 is at an advanced stage of evolution for human infection, while the human immune system is able to produce optimized antibodies. This discovery, unfortunately, implies the vulnerability of current vaccines and antibody drugs to new mutations. Our predictions were validated by  comparison with more than 1,400 deep mutations on the S protein RBD. Our results show the urgent need to develop new mutation-resistant vaccines and antibodies and to prepare for seasonal vaccinations.

\end{abstract}


\pagenumbering{roman}
\begin{verbatim}
\end{verbatim}
  \newpage


\setcounter{page}{1}
\renewcommand{\thepage}{{\arabic{page}}}
\section{Introduction}

The expeditious spread of coronavirus disease 2019 (COVID-19) pandemic caused by severe acute respiratory syndrome coronavirus 2 (SARS-CoV-2) has led to { 95,932,739 confirmed cases and 2,054,853 fatalities as of January 20, 2021.} In the 21st century, three major outbreaks of deadly pneumonia are caused by $\beta$-coronaviruses: SARS-CoV (2002), Middle East respiratory syndrome coronavirus (MERS-CoV) (2012), and SARS-CoV-2 (2019) \cite{lu2020genomic}. Similar to SARS-CoV and MERS-CoV, SARS-CoV-2  causes respiratory infections, and the transmission of viruses occurs among family members or in healthcare settings at the early stages of the outbreak. However, SARS-CoV-2 has an unprecedented scale of infection. Considering the high infection rate, high prevalence rate, long incubation period \cite{shin2020covid}, asymptomatic transmission \cite{day2020covid,long2020clinical}, and potential seasonal pattern \cite{kissler2020projecting} of COVID-19, the development of specific antiviral drugs, antibody therapies, and effective vaccines is of paramount importance. Traditional drug discovery takes more than ten years, on average, to bring a new drug    to  the market \cite{12yeartrip}. However, developing potent SARS-CoV-2 specified antibodies and vaccines is a relatively more efficient and less time-consuming strategy to combat COVID-19 for the ongoing pandemic \cite{bloch2020deployment}. Antibody therapies and vaccines depend on the host immune system. Recently studies have been working on the host-pathogen interaction, host immune responses, and the pathogen immune evasion strategies \cite{prompetchara2020immune, wu2020neutralizing, li2020orf6, tufan2020covid, liang2020highlight, catanzaro2020immune}, which provide insight into understanding the mechanism of antibody therapies and vaccine development. 

The immune system is a host defense system that protects the host from pathogenic microbes, eliminates toxic or allergenic substances, and responds to an invading pathogen \cite{chaplin2010overview}. It has innate immune system and  adaptive immune system as two major subsystems. The innate system provides an immediate but non-specific response, whereas the adaptive immune system provides a highly specific and effective immune response. Once the pathogen breaches the first physical barriers, such as epithelial cell layers, secreted mucus layer, mucous membranes, the innate system will be triggered to identify pathogens by pattern recognition receptors (PRRs), which is expressed on dendritic cells, macrophages, or neutrophils \cite{kumar2011pathogen}. Specifically, PPRs identify pathogen-associated molecular patterns (PAMPs) located on pathogens and then activate complex signaling pathways that introduce inflammatory responses mediated by various cytokines and chemokines, which promote the eradication of the pathogen \cite{takeuchi2010pattern,kumar2009pathogen}. Notably, the transmission of SARS-CoV-2 even occurs in asymptomatic infected individuals, which may delay the early response of the innate immune response \cite{prompetchara2020immune}. Another important line of host defense is the adaptive immune system. B lymphocytes (B cells) and T lymphocytes (T cells) are special types of leukocytes that are the acknowledged cellular pillars of the adaptive immune system \cite{pancer2006evolution}. Two major subtypes of T cells are involved in the cell-mediated immune response: killer T cells (CD8+ T cells) and helper T cells (CD4+ cells). The killer T cells eradicate cells  invaded by pathogens with the help of major histocompatibility complex (MHC) class I. MHC class I molecules are expressed on the surface of all nucleated cells \cite{hewitt2003mhc}. The nucleated cells will firstly degrade foreign proteins via antigen processing when viruses infect them. Then, the peptide fragments will be presented by MHC Class I, which will activate killer T cells to eliminate these infected cells by releasing cytotoxins \cite{harty2000cd8+}. Similarly, helper T cells cooperate with MHC Class II, a type of MHC molecules that are constitutively expressed on antigen-presenting cells, such as macrophages, dendritic cells, monocytes, and B cells \cite{ting2002genetic}. Helper T cells express T cell receptors (TCR) to recognize antigen bound to MHC class II molecules. However, helper T cells do not have cytotoxic activity. Therefore, they can not kill infected cells directly. Instead, the activated helper T cells will release cytokines to enhance the microbicidal function of macrophages and the activity of killer T cells \cite{alberts2018molecular}. Notably, an unbalanced response can result in a ``cytokine storm," which is the main cause of the fatality of COVID-19 patients \cite{hu2020cytokine}. Correspondingly, a B cell involves in humoral immune response and identifies pathogens by binding to foreign antigens with its B cell receptors (BCRs) located on its surface. The antigens that are recognized by antibodies will be degraded to { peptides} in B cells and displayed by MHC class II molecules. As mentioned above, helper T cells can recognize the signal provided by MHC class II and upregulate the expression of CD40 ligand, which provides extra stimulation signals to activate antibody-producing B cells \cite{grewal1998cd40}, rendering millions of copies of antibodies (Ab) that recognize the specific antigen. Additionally, when the antigen first enters the body, the T cells and B cells will be activated, and some of them will be differentiated to long-lived memory cells, such as memory T cells and memory B cells. These long-lived memory cells will play a role in quickly and specifically recognizing and eliminating a specific antigen that encountered the host  and initiated a corresponding immune response in the future \cite{crotty2004immunological}. The vaccination mechanism is to stimulate the primary immune response of the human body, which will activate T cells and B cells to generate the antibodies and long-lived memory cells that prevent infectious diseases, which is one of the most effective and economical means for combating with COVID-19 at this stage.

As mentioned above, secreted by B cells of the adaptive immune system, antibodies can recognize and bind to specific antigens. Conventional antibodies (immunoglobulins) are Y-shaped molecules that have two light chains and two heavy chains \cite{putnam1979primary}. Each light chain is connected to the heavy chain via a disulfide bond, and heavy chains are connected through two disulfide bonds in the mid-region known as the hinge region. Each light and heavy chain contain two distinct regions:  constant regions (stem of the Y) and variable regions (``arms" of the Y) \cite{wang2007antibody}. An antibody binds the antigenic determinant (also called epitope) through the variable regions in the tips of heavy and light chains. There is an enormous amount of diversity in the variable regions. Therefore, different antibodies can recognize many different types of antigenic epitopes. To be specific, there are three complementarity determining regions (CDRs) that are arranged non-consecutively in the tips of each variable region. CDRs generate most of the diversities between antibodies, which determine the specificity of individuals of antibodies. In addition to conventional antibodies, camelids also produce heavy-chain-only antibodies (HCAbs). HCAbs, also referred to as nanobodies, or VHHs, contain a single variable domain (VHH) that makes up the equivalent antigen-binding fragment (Fab) of conventional immunoglobulin G (IgG) antibodies \cite{hamers1993naturally}. This single variable domain typically can acquire affinity and specificity for antigens comparable to conventional antibodies. Nanobodies can easily be constructed into multivalent formats and have higher thermal stability and chemostability than most antibodies do \cite{van1999comparison}. Another advantage of nanobodies is that they are less susceptible to steric hindrances than large conventional antibodies \cite{forsman2008llama}.

Considering the broad specificity of antibodies, seeking potential antibody therapies has become one of the most feasible strategies to fight against SARS-CoV-2.  In general, antibody therapy is a form of immunotherapy that uses monoclonal antibodies (mAb) to target pathogenic proteins. The binding of antibody and pathogenic antigen can facilitate   immune response, direct neutralization, radioactive treatment, the release of toxic agents,   and  cytokine {storm } inhibition (aka immune checkpoint therapy). The SARS-CoV-2 entry of a human cell facilitated by the process of a series of interactions between its spike (S) protein and the host receptor angiotensin-converting enzyme 2 (ACE2), primed by host transmembrane protease, serine 2 (TMPRSS2) \cite{hoffmann2020sars}. As such, most COVID-19 antibody therapeutic developments focus on the SARS-CoV-2 spike protein antibodies that were initially generated from patient immune response and T-cell pathway inhibitors that block T-cell responses. A large number of antibody therapeutic drugs are in clinical trials. { Fifty-five} S protein antibody structures are available in the Protein Data Bank (PDB), offering a great resource for mechanistic analysis and biophysical studies.

\begin{figure}[H]
    \centering
    \includegraphics[width=0.9\textwidth]{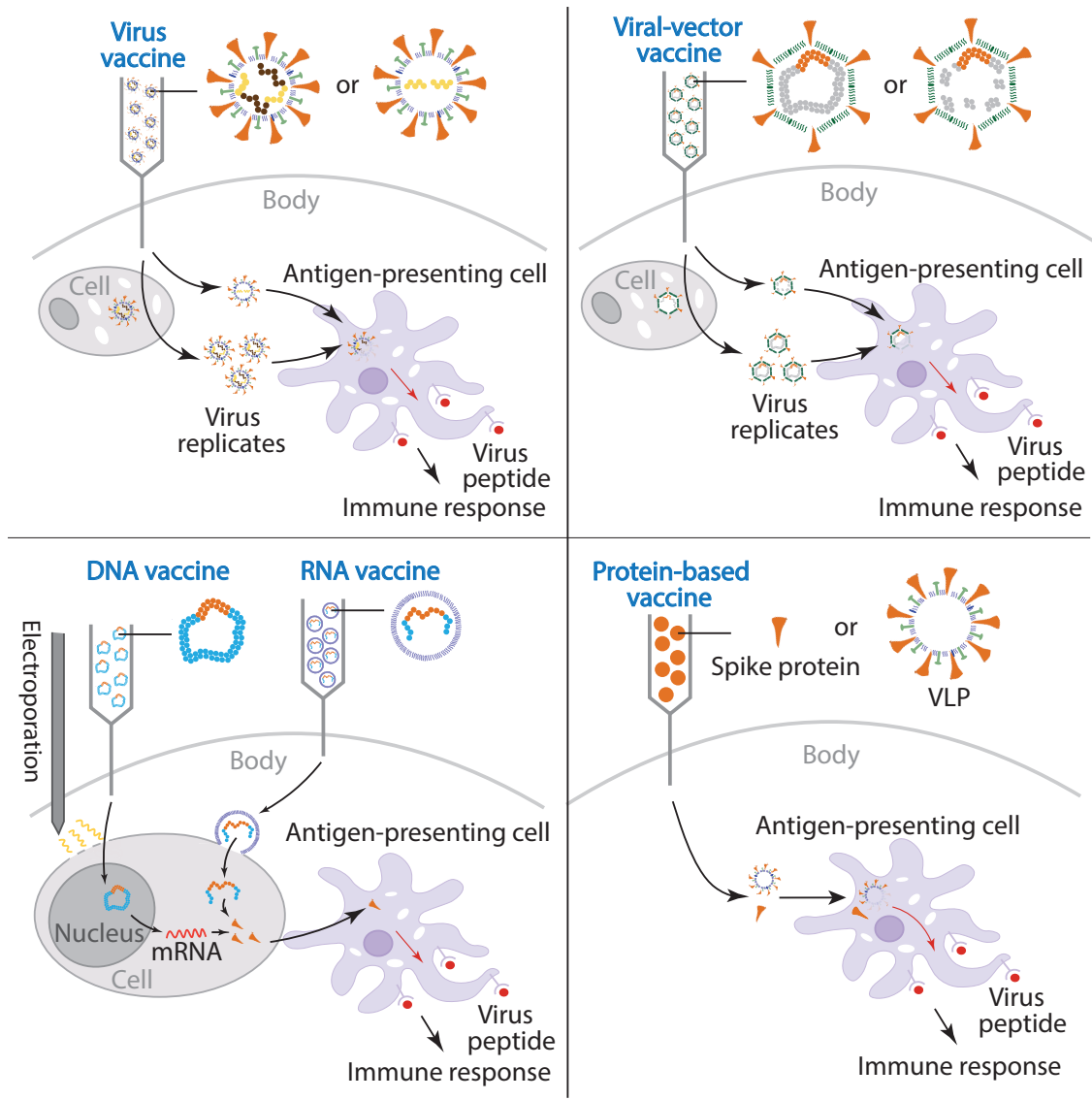}
    \caption{Illustration of four types of COVID-19 vaccines that are currently in the development. }
    \label{fig:vaccine}
\end{figure}

Currently, most antibody therapy developments focus on the use of antibodies isolated from patient convalescent plasma to directly neutralize SARS-CoV-2 \cite{cao2020covid,chen2020convalescent,shen2020treatment}, although there are efforts to alleviate cytokine storm. A more effective and economical means to fight against SARS-CoV-2 is   the  vaccine \cite{zhang2020progress}, which is the most anticipated approach for preventing the COVID-19 pandemic. A vaccine is designed to stimulate effective host immune responses and provide active acquired immunity by exploiting the body's immune system, including the production of antibodies, which is made of an antigenic agent that resembles a disease-causing microorganism, or surface protein, or genetic material that is needed to generate the surface protein. For SARS-CoV-2, the first choice of surface proteins is the spike protein. There are four types of COVID-19 vaccines, as shown in \autoref{fig:vaccine}. 1) Virus vaccines use the virus itself in a weakened or inactivated form. 2) Viral-vector vaccines are designed to genetically engineer a weakened virus, such as measles or adenovirus, to  produce coronavirus S proteins in the body. Both replicating and non-replicating viral-vector vaccines are being studied now. 3) Nucleic-acid vaccines use DNA or mRNA to produce SASR-CoV-2 S proteins inside host cells to stimulate the immune response. 4) Protein-based vaccines are designed to directly inject coronavirus proteins, such as S protein or membrane (M) protein, or their fragments, into the body.  Both protein subunits and viral-like particles (VLPs) are under development for COVID-19 \cite{callaway2020race}.  Among these technologies, nucleic-acid vaccines are safe and relatively easy to develop  \cite{callaway2020race}. However, they have not been approved for any human usage before. 

However, the general population's safety concerns are the major factors that hinder the rapid approval of vaccines and antibody therapies. A major potential challenge is an antibody-dependent enhancement, in which the binding of a virus to suboptimal antibodies enhances its entry into host cells. All vaccine and antibody therapeutic developments are currently based on the reference viral genome reported on January 5, 2020 \cite{wu2020new}. SARS-CoV-2 belongs to the coronaviridae family and the Nidovirales order, which has been shown to have a genetic proofreading mechanism regulated by non-structure protein 14 (NSP14) in synergy with NSP12, i.e.,  RNA-dependent RNA polymerase (RdRp) \cite{sevajol2014insights,ferron2018structural}. Therefore, SARS-CoV-2 has a higher fidelity in its transcription and replication process than other single-stranded RNA viruses, such as the flu virus and HIV. Even though the S protein of SARS-CoV-2 has been undergoing many mutations, as reported in \cite{wang2020decoding, wang2020decoding0}. As of { January 20, 2021}, a total of { 5,003 unique} mutations on the S protein has been detected on { 203,346} complete SARS-CoV-2 genome sequences. {Among them, 462 mutations were on the receptor-binding domain (RBD), the most popular targets for antibodies and vaccines.} Therefore, it is of paramount importance to establish a reliable paradigm to predict and mitigate the impact of SARS-CoV-2 mutations on vaccines and antibody therapies. Moreover, the efficacy of a given COVID-19 vaccine depends on many factors, including SARS-CoV-2 biological properties associated with the vaccine, mutation impacts, vaccination schedule (dose and frequency), idiosyncratic response, assorted factors such as ethnicity, age, gender, or genetic predisposition. The effect of COVID-19 vaccination also depends on the fraction of the population who accept vaccines. It is essentially unknown at this moment how these factors will unfold for COVID-19 vaccines. 

It is no doubt that any preparation that leads to an improvement in the COVID-19 vaccination effect will be of tremendous significance to human health and the world economy. Therefore, in this work, we integrate genetic analysis and computational biophysics, including artificial intelligence (AI), as well as additional enhancement from advanced mathematics to predict and mitigate mutation threats to COVID-19 vaccines and antibody therapies.  We perform single nucleotide polymorphism (SNP) calling \cite{wang2020decoding0} to identify SARS-CoV-2 mutations. For mutations on the S protein, we analyze their mechanism \cite{wang2020host}, frequency, ratio, and secondary structural traits. We construct a library of {{56}} existing antibody structures {by January 1, 2021} from the  PDB and analyze their two-dimensional (2D) and three-dimensional (3D) characteristics. We further predict the mutation-induced binding free energy (BFE) changes of antibody and S protein complexes using a topology-based network tree (TopNetTree) \cite{wang2020topology}, which is a state-of-the-art model that integrates deep learning and algebraic topology \cite{carlsson2009topology,edelsbrunner2000topological,xia2014persistent}.  In this work, TopNetTree is trained with newly available deep mutation datasets on the S protein, ACE2, and some antibodies and its predictions are validated with thousands of experimental data points. Our studies indicate that most mutations will significantly disrupt the binding of essentially all known antibodies to the S protein. Therefore, vaccines and antibody drugs that were developed based on the early SARS-CoV-2 genome will be seriously compromised by mutations. Additionally, we show that most known mutations will strengthen the binding between the S protein and ACE2, which gives rise to more infectious variants.  Our studies also reveal that SARS-CoV-2 is at an advanced stage of evolution with respect to its ability to infect human. Although the human immune system is able to produce antibodies that are optimized respect to a pathogen, the antibodies, once produced, are very vulnerable to the attack of mutants.  


\section{Mutations on the spike protein}\label{sec:S protein}
\begin{figure}[ht!]
    \centering
    \includegraphics[width=1\textwidth]{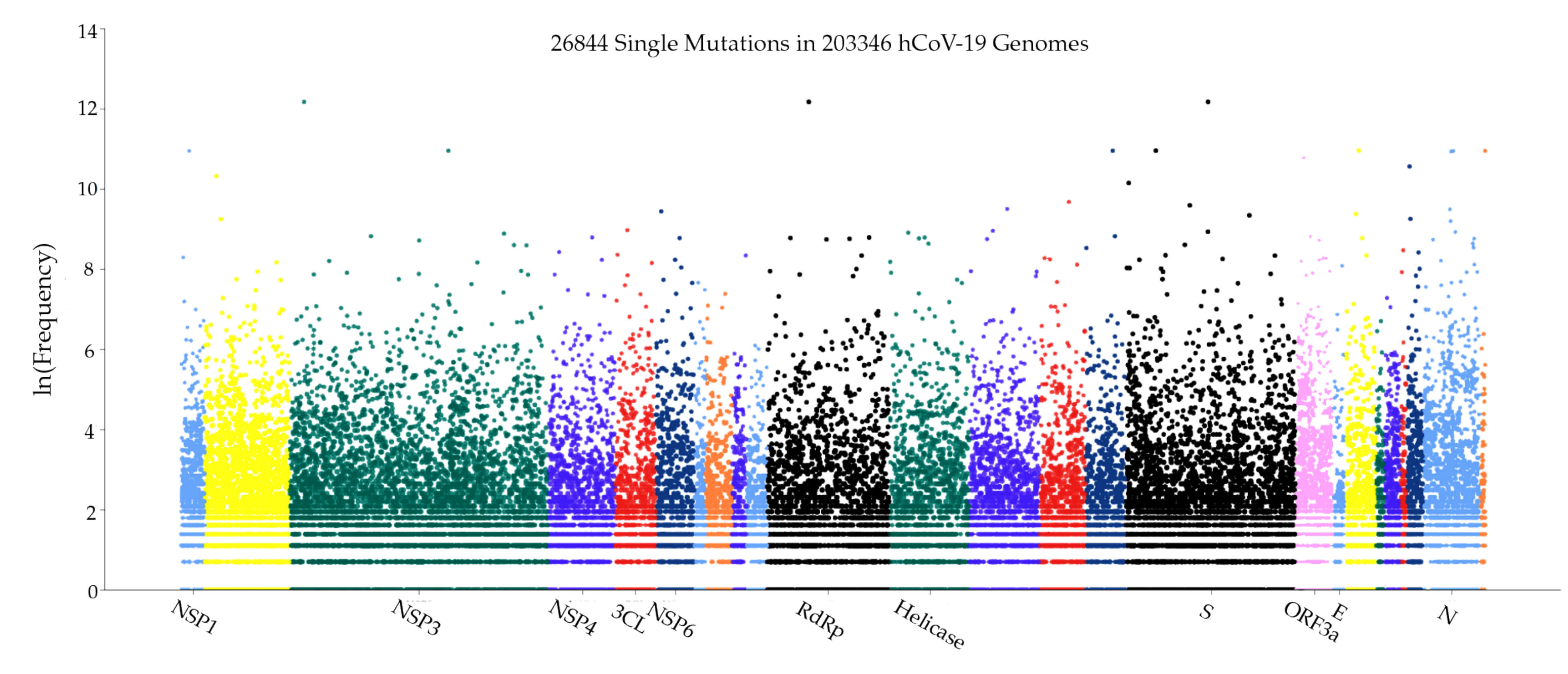}
    \caption{The distribution of genome-wide SARS-CoV-2 mutations on 26 proteins. The y-axis represents the natural log frequency for each mutation on a specific position of the complete SARS-CoV-2 genome. While only a few landmark positions are labeled with gene (protein) names, the relative positions of other genes(proteins) can be found in our Mutation Tracker \href{https://users.math.msu.edu/users/weig/SARS-CoV-2_Mutation_Tracker.html}{MutationTracker}(\url{https://users.math.msu.edu/users/weig/SARS-CoV-2_Mutation_Tracker.html})}
    \label{fig:Tracker}
\end{figure}

 As a fundamental biological process, mutagenesis changes the organism's genetic information and servers as a primary source for many kinds of cancer and heritable diseases, which is a driving force for evolution \cite{kucukkal2015structural,yue2005loss}. Generally speaking, virus mutations are introduced by natural selection, replication mechanism, cellular environment, polymerase fidelity, gene editing, random genetic drift, gene editing, recent epidemiology features, host immune responses, etc \cite{sanjuan2016mechanisms,grubaugh2020making}. Notably, understanding how mutations have changed the SARS-CoV-2 structure, function, infectivity, activity, and virulence is of great importance for coming up with life-saving strategies in virus control, containment, prevention, and medication, especially in the antibodies and vaccines development. Genome sequencing, SNP calling, and phenotyping provide an efficient means to parse mutations from a large number of viral samples \cite{wang2020decoding}(see the Supporting material (S1)). In this work, we retrieved  { more than  200,000} complete SARS-CoV-2 genome sequences from the GISAID database \cite{shu2017gisaid} and created a real-time interactive \href{https://users.math.msu.edu/users/weig/SARS-CoV-2_Mutation_Tracker.html}{SARS-CoV-2 Mutation Tracker} to report   {  more than 26,000} unique single mutations along with its mutation frequency on SARS-CoV-2 as of { January 20, 2021}. \autoref{fig:Tracker} is a screenshot of our online Mutation Tracker. It describes the distribution of mutations on the complete coding region of SARS-CoV-2. The $y$-axis shows the natural log frequency for each mutation at a specific position. A reader can download the detailed mutation SNP information from our Mutation Tracker website.

As mentioned before, the S protein has become the first choice for antibody and vaccine development. Among { 203,346} complete genome sequences, { 5,003} unique single mutations are detected on the S protein. The number of unique mutations (N$_{\text{U}}$) is determined by counting the same type of mutations in different genome isolates only once, whereas the number of non-unique mutations (N$_{\text{NU}}$, i.e., frequency) is calculated by counting the same type of mutations in different genome isolates repeatedly. \autoref{tab:gene} lists the distribution of 12 SNP types among unique and non-unique mutations on the S protein of SARS-CoV-2 worldwide. It can be seen that C$>$T and A$>$G are the two dominated SNP types, which may be due to the innate host immune response via APOBEC and ADAR gene editing \cite{wang2020host}.     

\begin{table}[ht!]
    \centering
    \setlength\tabcolsep{1pt}
	\captionsetup{margin=0.1cm}
	\caption{ { The distribution of 12 SNP types among 5,003 unique mutations and 467,604} non-unique mutations on the S gene of SARS-CoV-2 worldwide. N$_{\text{U}}$ is the number of unique mutations and N$_{\text{NU}}$ is the number of non-unique mutations. R$_{\text{U}}$ and R$_{\text{NU}}$ represent the ratios of 12 SNP types among unique and non-unique mutations.}
    \label{tab:gene}
    {
    \begin{tabular}{clcccc|clccccccccc}
    \toprule
    SNP Type & Mutation Type & N$_{\text{U}}$ & N$_{\text{NU}}$ & R$_{\text{U}}$  & R$_{\text{NU}}$ & SNP Type & Mutation Type  & N$_{\text{U}}$ & N$_{\text{NU}}$ & R$_{\text{U}}$  & R$_{\text{NU}}$\\
    \midrule
    A$>$T & Transversion & 454 & 5236 & 9.07\% & 1.12\%  & C$>$T & Transition    & 542 & 158898 & 10.83\% & 33.98\% \\
    A$>$C & Transversion & 341 & 2571 & 6.82\% & 0.55\% & C$>$A & Transversion  & 313 & 10301 & 6.26\% & 2.20\%\\
    A$>$G & Transition   & 700 & 199015 & 13.99\% & 42.56\% & C$>$G & Transversion  & 156 & 968 & 3.12\% & 0.21\%\\
    T$>$A & Transversion & 356 & 1614 & 7.12\% & 0.35\%  & G$>$T & Transversion  & 435 & 34421 & 8.69\% & 7.36\%\\
    T$>$C & Transition   & 779 & 19313 & 15.57\% & 4.13\%  & G$>$C & Transversion  & 225 & 6090 & 4.50\% & 1.30\%\\
    T$>$G & Transversion & 277 & 1940 & 5.54\% & 0.41\%  & G$>$A & Transition    & 425 & 27237 & 8.49\% & 5.82\%\\
    \bottomrule
    \end{tabular}
    }
\end{table}

Moreover, { 144} non-degenerated mutations occurred on the S protein RBD, which are relevant to the binding of SARS-CoV-2 S protein and most antibodies as well as ACE2. Additionally, { 218} mutations { that} occurred on the S protein NTD (residue id: 14 to 226) are relevant to the binding of another { two antibodies (4A8, FC05)} and SARS-CoV-2 S protein.

Furthermore, since antibody CDRs are random coils, the complementary antigen-binding domains must involve random coils as well. \autoref{tab:pro} lists the statistics of non-degenerate mutations on the secondary structures of SARS-CoV-2 S protein. Here, the secondary structures are mostly extracted from the crystal structure of 7C2L\cite{chi2020neutralizing}, and the missing residues are predicted by RaptorX-Property \cite{wang2016raptorx}. We can see that for both unique and non-unique cases, the average mutation rates on the random coils of the S protein have the highest values. Particularly, the 23403A$>$G-(D614G) mutation on the random coils has the highest frequency of { 192284}. If we do not consider the 23403A$>$G-(D614G) mutations, then the unique and non-unique average rates on the random coils of S protein still have the highest values { (2.81 and 212.01)}, indicating that mutations are more likely to occur on the random coils. Consequently, the natural selection of mutations may tend to disrupt antibodies.

\begin{table}[ht!]
    \centering
    \setlength\tabcolsep{11pt} 
	\captionsetup{margin=0.1cm}
	\caption{The statistics of non-degenerate mutations on the secondary structure of SARS-CoV-2 S protein. The unique and non-unique mutations are considered in the calculation. N$_{\text{U}}$, N$_{\text{NU}}$, AR$_{\text{U}}$, AR$_{\text{NU}}$ represent the number of unique mutations, the number of non-unique mutations, the average rate of unique mutations, and the average rate of non-unique mutations on the secondary structure of S protein, respectively. Here, the secondary structure is mostly extracted from the crystal structure of 7C2L, the missing residues are predicted by RaptorX-Property. }
    \label{tab:pro}
    { \begin{tabular}{c|ccccc}
    \toprule
    Secondary structure & Length & N$_{\text{U}}$ & N$_{\text{NU}}$  & AR$_{\text{U}}$  & AR$_{\text{NU}}$ \\
    \midrule
    Helix        & 249   & 516  & 9535    & 2.07 & 38.29 \\
    Sheet        & 276   & 711  & 20422    & 2.58 & 73.99 \\
    Random coils & 748   & 2100  & 350659   & \bf{2.81} & \bf{468.80} \\
    Whole Spike        & 1273  & 3327 & 380616   & 2.61 & 298.99 \\
    \bottomrule
    \end{tabular}
    }
\end{table}

\section{SARS-CoV-2 antibodies}

{ In this work, we consider 56 3D structures available from the PDB (\url{https://www.rcsb.org}) before January 1, 2021. These 56 structures include 51 structures of antibodies binding to S protein RBD, 4 structures of antibodies  having binding domains outside the S protein RBD, and an ACE2-S protein complex.     Among four structures having binding domains outside the RBD, there are three distinct antibodies not binding to the RBD, namely  4A8 \cite{chi2020neutralizing}, FC05 \cite{wang2021structure}, and 2G12 \cite{acharya2020glycan}. This is because that FC05 has two sets of structures (PBD IDs 7CWU and 7CWS) that differ from each other by their components on the RBD (i.e., H014 and P17). Some antibodies are given as combinations of other unique ones. Among 51 antibodies on the RBD, there are only 42 unique ones, including MR17-K99Y as a mutant of MR17  \cite{li2020potent}.
}
 
\begin{figure}[h!]
	\centering
	\includegraphics[width=0.8\textwidth]{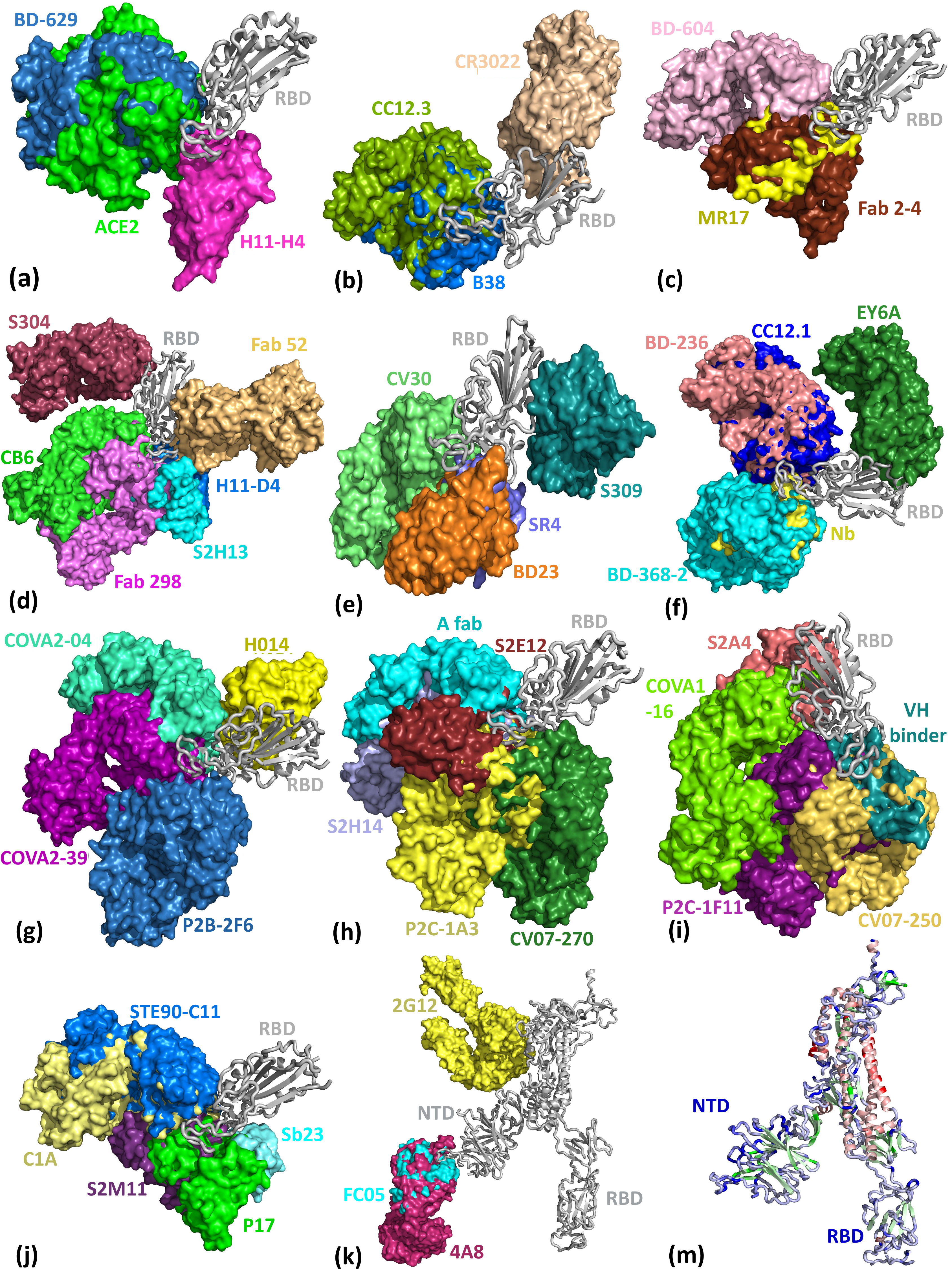}
	\caption{{Aligned structures of 46 complexes of the S protein and ACE2 and single antibodies.} (a)-(j) The 3D alignment of the available unique 3D structures of SARS-CoV-2 S protein RBD in binding complexes with { 42} antibodies (MR17-K99Y is excluded because its binding mode is the same as that of MR17). (k) The 3D alignment of the three antibodies binding outside RBD. (m) The 3D structure of S protein RBD. The red, green, and blue colors represent for helix, sheet, and random coils of RBD, respectively. The darker color represents the higher mutation frequency on a specific residue. {The structures are 
	(a) ACE2 (6M0J)\cite{lan2020structure}, BD-629 (7CH5), H11-H4 (6ZBP); 
	(b) CC12.3 (6XC4)\cite{yuan2020structural}, B38 (7BZ5)\cite{wu2020noncompeting}, CR3022 (6XC3)\cite{yuan2020structural};  
	(c) BD-604 (7CH4), MR17 (7C8W) \cite{li2020potent}, Fab 2-4 (6XEY)\cite{li2020potent}; 
	(d) S304 (7JW0)\cite{piccoli2020mapping}, CB6 (7C01)\cite{shi2020human}, Fab 52 (7K9Z) \cite{rujas2020multivalency}, S2H13 (7JV6)\cite{piccoli2020mapping}, H11-D4 (6YZ5) \cite{huo2020structural}, Fab 298 (7K9Z)\cite{rujas2020multivalency}; 
	(e) CV30 (6XE1) \cite{hurlburt2020structural}, BD23 (7BYR)\cite{cao2020potent}, SR4 (7C8V) \cite{li2020potent}, S309 (6WPS)\cite{pinto2020structural}; 
	(f) CC12.1 (6XC2) \cite{yuan2020structural}, EY6A (6ZCZ)\cite{zhou2020structural}, BD-236 and nanobody (Nb) (7CHE)\cite{du2020structurally}, BD-368-2 (7CHH)\cite{du2020structurally}; 
	(g) H014 (7CAH)\cite{lv2020structural}, COVA2-04 (7JMO)\cite{wu2020alternative}, COVA2-39 (7JMP)\cite{wu2020alternative}, P2B-2F6 (7BWJ)\cite{ju2020human}; 
	(h) P2C-1A3 (7CDJ),  CV07-270 (6XKP)\cite{kreye2020therapeutic}, S2H14 (7JX3)\cite{piccoli2020mapping}, A fab (7CJF), S2E12 (7K45)\cite{tortorici2020ultrapotent};
	(i) CV07-250 (6XKQ)\cite{kreye2020therapeutic}, P2C-1F11 (7CDI), VH binder (7JWB)\cite{bracken2021bi}, S2A4 (7JVA)\cite{piccoli2020mapping}, COVA1-16 (7JMW) \cite{liu2020cross}, 
	(j) C1A (7KFV)\cite{clark2020molecular}, STE90-C11 (7B3O)\cite{bertoglio2020sars}, Sb23 (7A29)\cite{custodio2020selection}, S2M11 (7K43)\cite{tortorici2020ultrapotent}, P17 (7CWM) \cite{yao2021rational};
	(k) 4A8 (7C2L)\cite{chi2020neutralizing}, FC05 (7CWU)\cite{wang2021structure}, and 2G12 (7L06)\cite{acharya2020glycan}.}}
	\label{fig:antibodies}
\end{figure}

\subsection{3D antibody structure alignment on the S protein}

{We present the 3D alignment of 45 structures of SARS-CoV-2 S protein with ACE2 and antibodies (excluding the mutant MR17-K99Y of MR17) in Figure \ref{fig:antibodies}. }
 ACE2 in Figure \ref{fig:antibodies} (a) is a reference. {Figures \ref{fig:antibodies} (a)-(j) list 42 single antibodies binding to the RBD, and Figure \ref{fig:antibodies} (k) includes the other 3 alignments of 4A8, FC05, and 2G12 whose binding domains are outside the RBD. Figure \ref{fig:antibodies} (m) presents a 3D structure of a single chain of S protein.}  The PDB IDs of these complexes can be found in Figure \ref{fig:sars-epitope-2d-align}.

Figure \ref{fig:antibodies} reveals, { except for Fab 52\cite{rujas2020multivalency}, S309\cite{lan2020structure}, CR3022\cite{huo2020structural}, EY6A\cite{zhou2020structural},  4A8\cite{chi2020neutralizing}, FC05 \cite{wang2021structure}, and 2G12 \cite{acharya2020glycan}, all the other 38 antibodies have their binding sites spatially clashing with that of ACE2. Notably, the paratope of H014\cite{lv2020structural} and S304 \cite{piccoli2020mapping} do not overlap with that of ACE2 directly, but in terms of 3D structures, their binding sites still overlap. This suggests that the bindings of 39 antibodies are in direct competition with that of ACE2}. Theoretically, this direct competition reduces the viral infection rate. For such an antibody with strong binding ability, it will directly neutralize SARS-CoV-2 without the need {for} antibody-dependent cell cytotoxicity (ADCC), antibody-dependent cellular phagocytosis (ADCP), or other immune mechanisms.

The paratopes of S309, Fab 52, CR3022, and EY6A on the RBD are away from that of ACE2, leading to the absence of binding competition \cite{pinto2020structural, tian2020potent,zhou2020structural}. One study shows that the ADCC and ADCP mechanisms contribute to the viral control conducted by S309 in infected individuals \cite{pinto2020structural}. { For Fab 52, it was suggested that its mechanism could involve S protein destabilization \cite{rujas2020multivalency}. For CR3022, one research indicates that it neutralizes the virus in a synergistic fashion \cite{ter2006human}.}  For EY6A, the hypothesis is that { glycosylation  of  ACE2  accounts  for  at  least part of the observed crosstalk between ACE2 and EY6A.}   \cite{zhou2020structural}. More radical examples are 4A8, { FC05, and 2G12.} 4A8 binds to NTD of the S protein (Figure \ref{fig:antibodies}(h)), which is quite far from the RBD. It is speculated that 4A8 may neutralize SARS-CoV-2 by restraining the conformational changes of the S protein, which is very important for the SARS-CoV-2 cell entry  \cite{chi2020neutralizing}. { FC05 is combined with P17 or H014 to form a cocktail \cite{wang2021structure}. 2G12 binds to the S protein S2 domain \cite{acharya2020glycan}.}  Any antibody or drug that can inhibit serine protease TMPRSS2 priming of the S protein priming can effectively stop the viral cell entry \cite{hoffmann2020sars}.

\subsection{2D residue contacts between antibodies and the S protein RBD} \label{epitode_alignment}

\begin{figure}[ht!]
	\centering
	\includegraphics[width=1.0\textwidth]{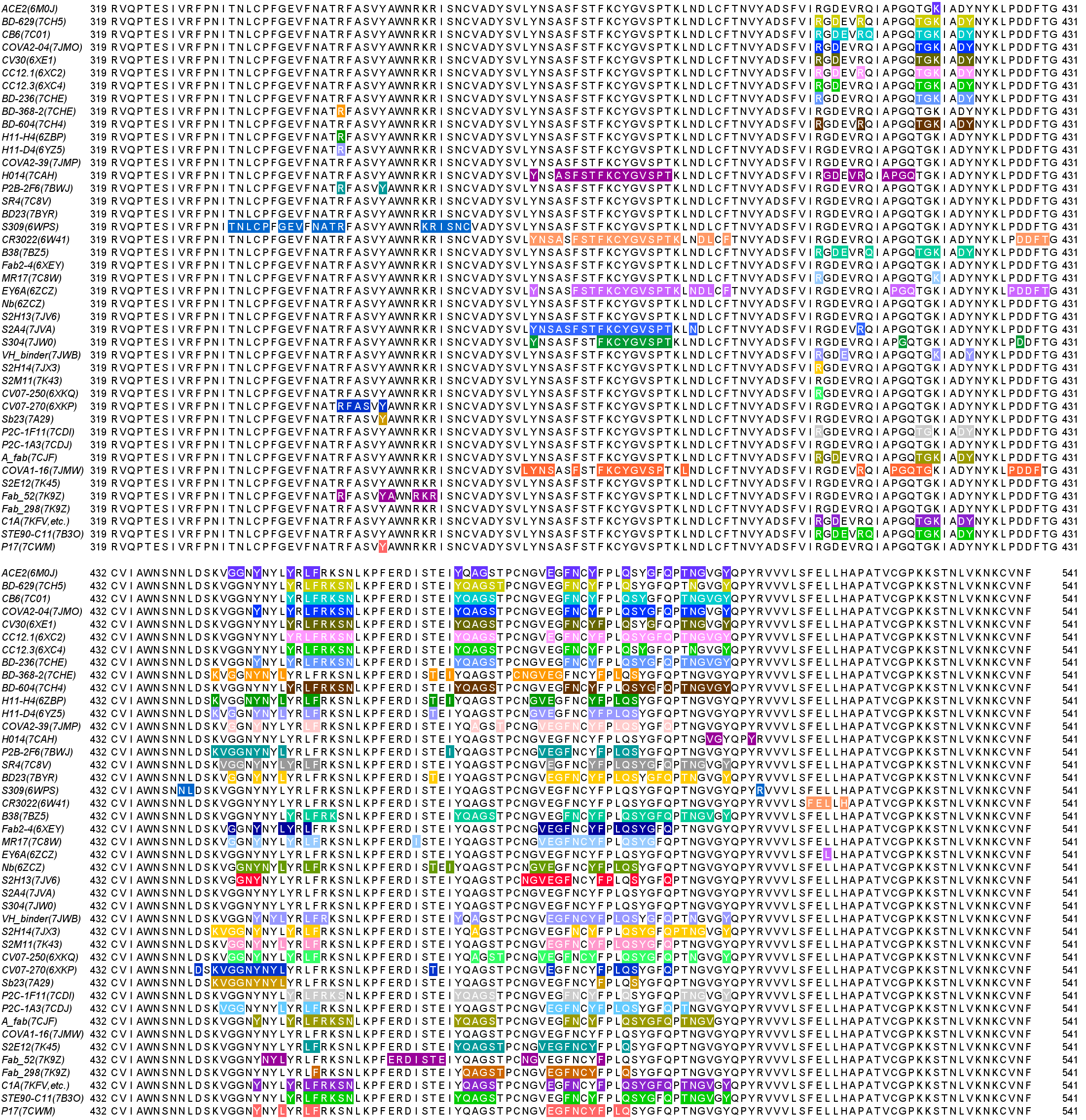}
	\caption{Illustration of the contact positions of  antibody and ACE2   paratope  with SARS-CoV-2 S protein RBDs on RBD 2D sequences. The corresponding PDB IDs   are given in parentheses.}
	\label{fig:sars-epitope-2d-align}
\end{figure}

Figure \ref{fig:antibodies} provides a visual illustration of antibody and ACE2 competitions. It remains to know in the residue detail what has happened to these competitions. To better understand the antibody and S protein interactions, we study the residue contacts between antibodies and the S protein.  We include the ACE2 as a reference but excluding antibodies MR17-K99Y as well as 4A8, FC05, and 2G12 that bind to other domains.

In Figure \ref{fig:sars-epitope-2d-align}, the paratopes  of { 42 individual antibodies (excluding MR17-K99Y)} and ACE2 were aligned on the S protein RBD 2D sequence, and their contact regions are highlighted. From the figure, one can see that, except for { Fab 52, S309, CR3022, EY6A, H014, and S304,} all the other { 36} antibodies have their antigenic epitopes overlapping with the ACE2, especially on the residues from 486 to 505 of the RBD.  {Although}, the paratope of H014 and S304 do not overlap with that of ACE2 directly,  {their binding sites still overlap in 3D structures}. Therefore, these { 38} antibodies competitively bind against ACE2 as revealed in Figure \ref{fig:antibodies}. 

\subsection{Antibody sequence alignment and similarity analysis} \label{seq_alignment}

 The next question is whether there is any connection or similarity between the antibody paratopes in our library, particularly for those antibodies that share the same binding sites. To better understand this perspective,  we carry out multiple sequence alignment (MSA) to further study the similarity and difference among existing antibodies.  Many antibodies are very similar to each other and can be classified into several clusters by CD-HIT suite\cite{huang2010cd}. { The first and largest cluster includes COVA2-04, CC12.1, BD-236, BD-604, B38, EY6A, S304, P2C-1A3, A fab, C1A, STE90-C11, and CB6. Their identity scores to CB6 are 90.48 \%, 94.74 \%, 93.59 \%, 93.35 \%, 94.77 \%, 92.52 \%, 90.62 \%, 90.51 \%, 91.18 \%, 94.08 \%, and 93.00 \%, respectively. The second cluster contains BD-629, CC12.3, P2C-1F11, and CV30. Their identity scores to CV30 are 95.41 \%, 96.32 \%, and 97.68 \%, respectively. The third cluster has CV07-270 and COVA2-39, the pairwise identity score is 90.18 \%. The fourth cluster is composed of H11-H4, H11-D4, and Nb, and their identity scores to Nb are 99.25 \% and 95.52 \%, respectively. They are all nanobodies. The fifth cluster has Fab 298 and COVA1-16, the pairwise identity score is 90.80 \%. Their alignment plots are in the Supporting material (Figures S1-S5).}

The above similarity indicates that the adaptive immune systems of individuals have a common way to generate antibodies. On the other hand, the existence of five distinct clusters, as well as antibodies { 4A8, FC05, and 2G12} suggests the diversity in the immune response. Note that we have also included ACE2 in our MSA as a reference, but none of the existing antibodies is similar to ACE2 because they were created from entirely different mechanisms.

\section{Mutation impacts on SARS-CoV-2 antibodies}

To investigate the influences of existing S protein mutations on the binding free energy (BFE) of S protein and antibodies, we consider {462} mutations { that} occurred on the S protein RBD, which are relevant to the binding of SARS-CoV-2 S protein and antibodies as well as ACE2. Additionally, {540} mutations occurred on the NTD of the S protein (residue id: 14 to 226) which are relevant to the binding of SARS-COV-2 S protein and antibody 4A8 (PDB: 7C2L). We predict the free energy changes following existing mutations using our TopNetTree model \cite{wang2020topology}.  {The mutations on the RBD are considered for the predictions of BFE changes.} Our predictions are built from the X-ray crystal structure of SARS-CoV-2 S protein and ACE2 (PDB 6M0J) \cite{lan2020structure}, and various antibodies (PDBs 6WPS\cite{pinto2020structural}, 6XC2\cite{yuan2020structural}, 6XC3\cite{yuan2020structural}, 6XC4\cite{yuan2020structural}, 6XC7\cite{yuan2020structural},  6XE1\cite{hurlburt2020structural}, 6XEY\cite{liu2020potent}, {6XKP}\cite{kreye2020therapeutic}, {6XKQ}\cite{kreye2020therapeutic}, 6YLA\cite{huo2020structural}, 6YZ5, 6Z2M, 6ZBP, 6ZCZ\cite{zhou2020structural}, 6ZER\cite{zhou2020structural}, {7A29\cite{custodio2020selection}, 7B3O,} 7BWJ\cite{ju2020human}, 7BYR\cite{cao2020potent}, 7BZ5\cite{wu2020noncompeting}, 7C01\cite{shi2020human}, 7C2L\cite{chi2020neutralizing}, 7C8V\cite{li2020potent}, 7C8W\cite{li2020potent}, 7CAH\cite{lv2020structural}, {7CAH\cite{lv2020structural},} 7CAN\cite{li2020potent}, {7CDI, 7CDJ,} 7CH4\cite{du2020structurally}, 7CH5\cite{du2020structurally}, 7CHB\cite{du2020structurally}, 7CHE\cite{du2020structurally}, 7CHF\cite{du2020structurally}, 7CHH\cite{du2020structurally}, {7CJF}, {7CWM}\cite{yao2021rational}, {7CWN}\cite{yao2021rational} 7JMO\cite{wu2020alternative}, 7JMP\cite{wu2020alternative}, {7JMW}\cite{liu2020cross}, {7JV6}\cite{piccoli2020mapping}, {7JVA}\cite{piccoli2020mapping}, {7JVC}\cite{piccoli2020mapping}, {7JW0}\cite{piccoli2020mapping}, {7JWB}\cite{bracken2021bi}, {7JX3}\cite{piccoli2020mapping}, {7K43}\cite{tortorici2020ultrapotent}, {7K45}\cite{tortorici2020ultrapotent}, {7K9Z}\cite{rujas2020multivalency}, {7KFV}\cite{clark2020molecular}, {7KFW}\cite{clark2020molecular}, {7KFX}\cite{clark2020molecular}, and {7KFY}\cite{clark2020molecular}). The BFE change following mutation ($\Delta\Delta G$) is defined as the subtraction of the BFE of the mutant type from the BFE of the wild type, $\Delta\Delta G=\Delta G_{\rm W} - \Delta G_{\rm M}$ where $\Delta G_{\rm W}$ is the BFE of the wild type and $\Delta G_{\rm M}$ is the BFE of mutant.  Therefore, a negative BFE change means that the mutation decreases affinities, making the protein-protein interaction less stable. 

{
Four antibody-S protein complexes are examined in in this section.  Next, we present a library of mutation-induced BFE changes for all mutations and 51 antibodies, as well as ACE2. The statistical analysis of mutation impacts on antibodies is discussed.}

\subsection{Single antibody-S protein complex analysis}
\begin{figure}[ht]
	\centering
	\includegraphics[width=1\textwidth]{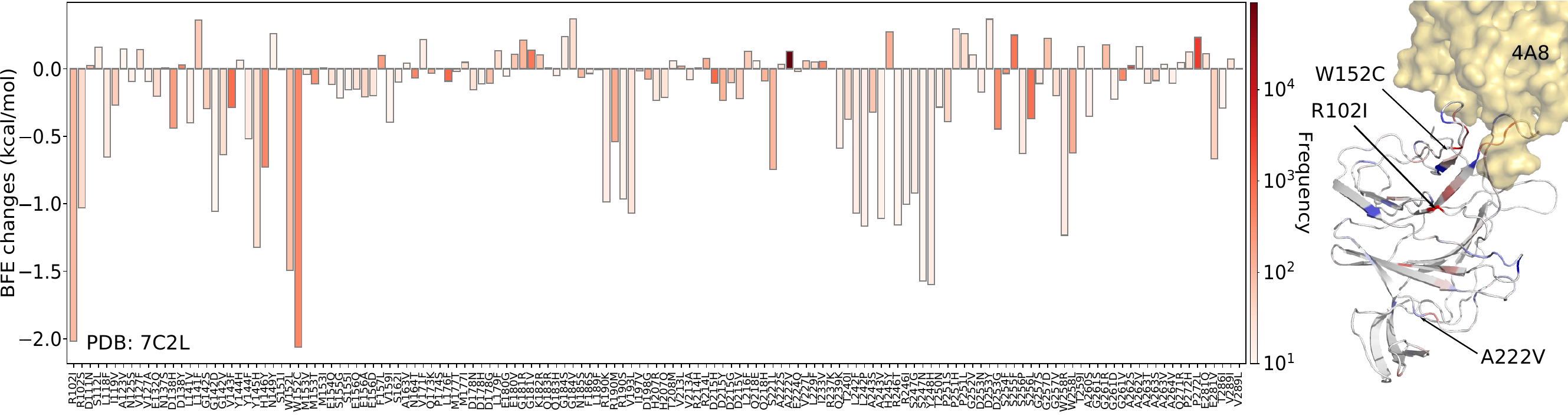}
	\caption{Illustration of SARS-CoV-2 mutation-induced binding free energy changes for the complexes of S protein and 4A8 (PDB: 7C2L). The blue color in the structure plot indicates a positive BFE change while the red color indicate a negative BFE change, and toning indicate the strength. Here, mutations {R102I, W152C, W152L, S247N, and Y248H} could potentially disrupt the binding of antibody 4A8 and S protein.
	}
	\label{fig:7C2L_4A8}
\end{figure}

{For four  antibody-S protein complexes, since there are too many mutations, we only consider those  mutations whose frequencies are greater than 10.}
We first present the BFE changes ($\Delta\Delta G$) of SARS-CoV-2 S protein binding domain with antibody 4A8 in Figure~\ref{fig:7C2L_4A8}, which is {one of the three complexes} that are not on the RBD in our collections of S protein and antibody complexes. {A total 141 of 540 mutations on residues ID from 14 to 226 have frequencies larger than 10.} Most mutations have small BFE changes {(from $-0.5$ kcal/mol to $0.5$ kcal/mol)} in their binding free energies, while { 28 mutations have negative BFE changes less than $-0.5$ kcal/mol}. Notably, {53 out of 141}  mutations on the binding domain have positive BFE changes, which means that the mutations increase affinities and would make the S protein-4A8 interactions more stable. However, the majority {(63\%)} of mutations have negative BFE changes, including high-frequency mutations, {R102I, and W152C with frequencies of 89 and 356, respectively. Since the largest positive and negative BFE changes are 0.37 and -2.06 (-3.1 if low frequency mutations are counted) kcal/mol, respectively, the prediction indicates that antibody 4A8 isolated from 10 convalescent patients at the early stage of the pandemic \cite{chi2020neutralizing} is an optimized product of the human immune system with respect to the original S protein.}  It is also noted that many mutations on the binding domain, such as {W152L, S247N, and Y248H}, have significant negative free energy changes. The mutations on the binding domain with large negative BFE changes reveal that the binding of antibody 4A8 and S protein will  be potentially  disrupted. 

\begin{figure}[ht]
	\centering
	\includegraphics[width=1\textwidth]{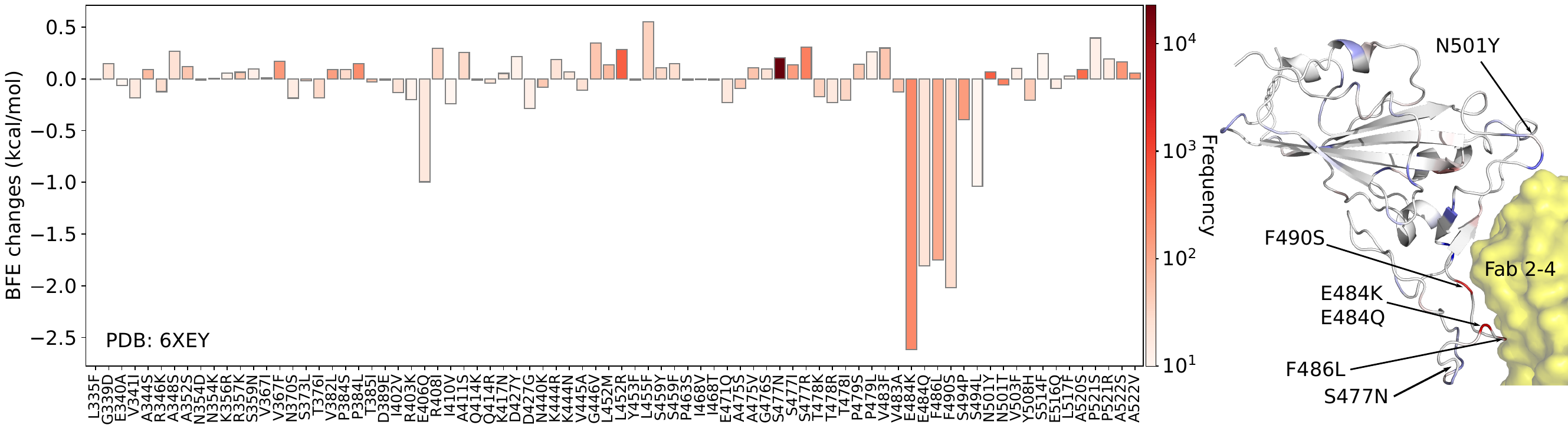}
	\caption{Illustration of SARS-CoV-2 mutation-induced binding free energy changes for the complexes of S protein and Fab 2-4 (PDB: 6XEY). The blue color in the structure plot indicates a positive BFE change while the red color indicate a negative BFE change, and toning indicate the strength. Here, mutations {E484K, E484Q, F486L, and F490S} could potentially disrupt the binding of antibody Fab 2-4 and the S protein.
	}
	\label{fig:6XEY_Fab_2-4}
\end{figure}
Next, we study the BFE changes ($\Delta\Delta G$) induced by {80} mutations on the SARS-CoV-2 S protein RBD for the antibody Fab 2-4 (PDB: 6XEY) in Figure~\ref{fig:6XEY_Fab_2-4}. Antibody Fab 2-4 shares a similar binding domain with ACE2 and thus is a potential candidate for the direct neutralization of SARS-CoV-2. Most mutations induce small changes in the binding free energies, while mutations, {E484K, E484Q, F486L, and F490S}, have large negative BFE changes. Overall, 38 out of 80 mutations on the RBD lead to negative BFE changes, which means 48\% of mutations will potentially weaken the binding between antibody Fab 2-4 and S protein. {For positive BFE changes, the largest value is only 0.55 kcal/mol and the average of positive BFE changes is 0.16 kcal/mol. }    {However, many mutations with negative BFE changes have   a very large magnitude, indicating that antibody Fab 2-4 was an immune product optimized with respect to the original un-mutated S protein. In general,  the mutations on S protein weaken the Fab 2-4 binding with S protein and make  it less competitive with ACE2 as most mutations strengthen the S protein and ACE2 binding.
It is interesting to note that mutation E484K is the so-called South Africa variant. It indeed has a strong vaccine-escape effect.     
}

\begin{figure}[ht]
	\centering
	\includegraphics[width=1\textwidth]{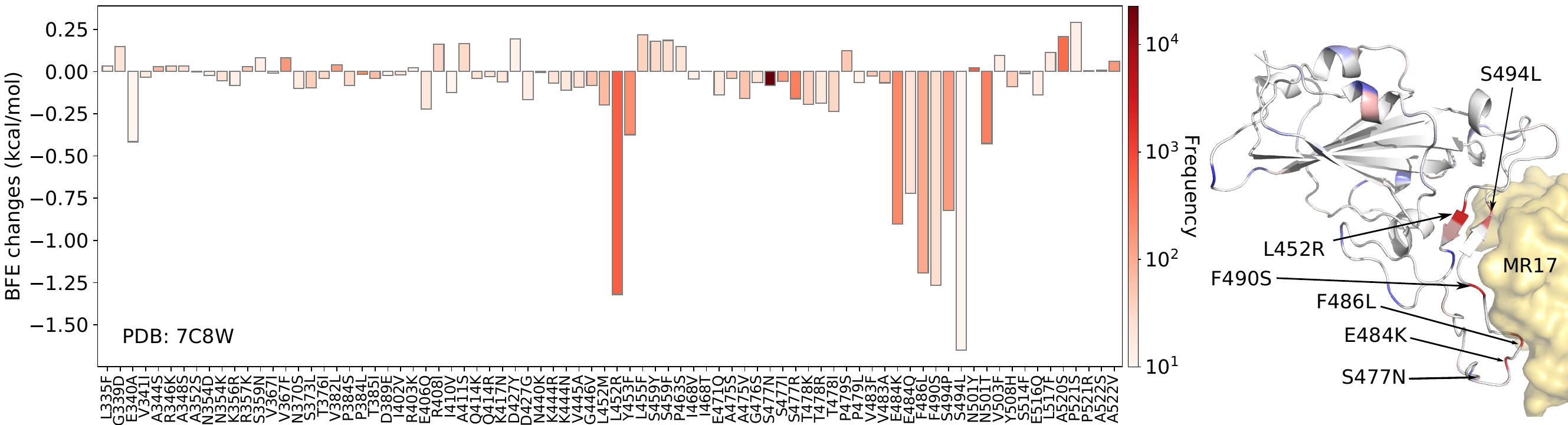}
	\caption{Illustration of SARS-CoV-2 mutation-induced binding free energy changes for the complexes of S protein and MR17 (PDB: 7C8W). Blue in the structure plot indicates a positive BFE change while red indicate a negative BFE change, and toning indicate the strength. Here, mutations {L452R, E484K, F486L, F490S, and S494L} could potentially disrupt the binding of antibody MR17 and the S protein.
	}
	\label{fig:7C8W_MR17}
\end{figure}
In Figure~\ref{fig:7C8W_MR17}, we illustrate the mutation-induced BFE changes for antibody MR17 (PDB: 7C8W), which shares the binding domain with ACE2 as well. One can notice that four mutations, {L452R, E484K, F486L, F490S, and S494L}, have BFE changes less than $-1$ kcal/mol {as well as high frequencies}. The rest mutations have a small magnitude of changes. {Twenty seven  out of 80 mutations have positive BFE changes with the largest value less than 0.25 kcal/mol.} 
{Our results indicate that antibody MR37 is likely to be isolated from patients at the early stage and thus, it was optimized based on an early version of SARS-CoV-2 virus. Mutations L452R, E484K, F486L, F490S, and S494L} will reduce its competitiveness with ACE2.

\begin{figure}[ht]
	\centering
	\includegraphics[width=1\textwidth]{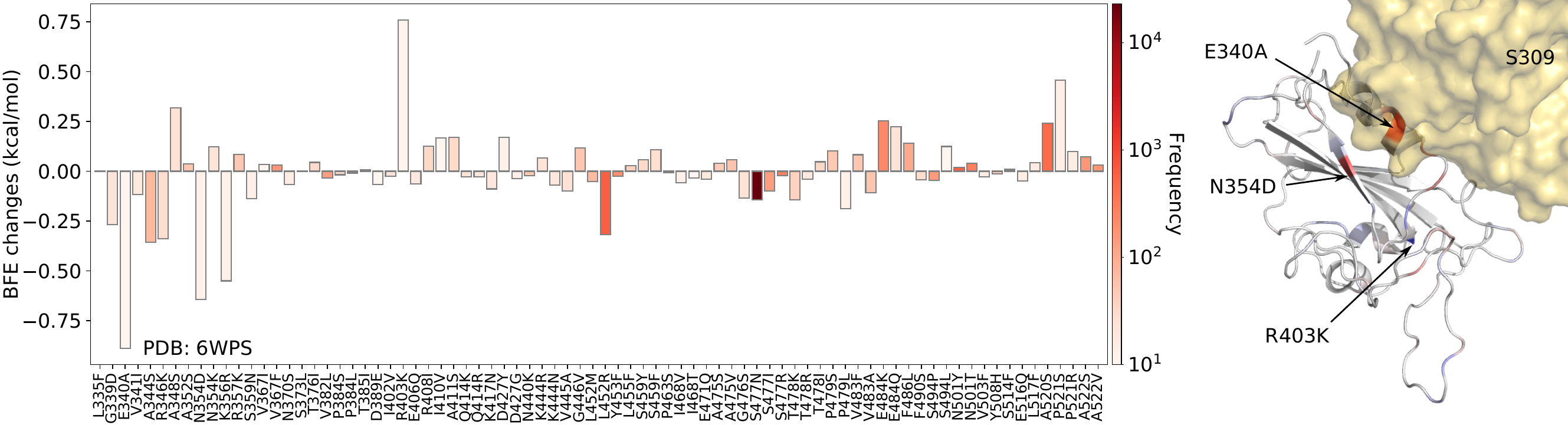}
	\caption{Illustration of SARS-CoV-2 mutation-induced binding free energy changes for the complexes of S protein and S309 (PDB: 6WPS). The blue color in the structure plot indicates a positive BFE change while the red color indicate a negative BFE change, and toning indicate the strength. Here, mutations {E340A, N354D, and K356R} could potentially weaken the binding of antibody S309 and the S protein.
	}
	\label{fig:6WPS_S309}
\end{figure}
Finally, we consider the BFE change predictions for antibody S309 and S protein complex, whose receptor binding motif (RBM) does not overlap with the RBM of ACE2 (see Fig. \ref{fig:antibodies}(e)). The BFE changes induced by 80 mutations are predicted. Among them, 38 changes are positive. Similar to the aforementioned antibodies, most of the mutations lead to small changes in their binding affinity magnitude but  three mutations, {E340A, N354D, and K356R, induce moderate} negative changes. {Interestingly, 
all the 80 RBD mutations do not have a major impact on S309. Although mutation R403K might disrupt S309, it does weaken  many other antibody bindings with the S protein.}  
 While antibodies play a variety of functions in the human immune system, such as neutralization of infection, phagocytosis, antibody-dependent cellular cytotoxicity, etc., their binding with antigens is crucial for these functions.  Our analysis of BFE changes following mutations on the S protein suggests that some antibodies will be less affected by mutations, which is important for developing vaccine and antibody therapies. 

\subsection{Mutation impact library}

\begin{figure}[ht!]
\centering
\includegraphics[width=1\textwidth]{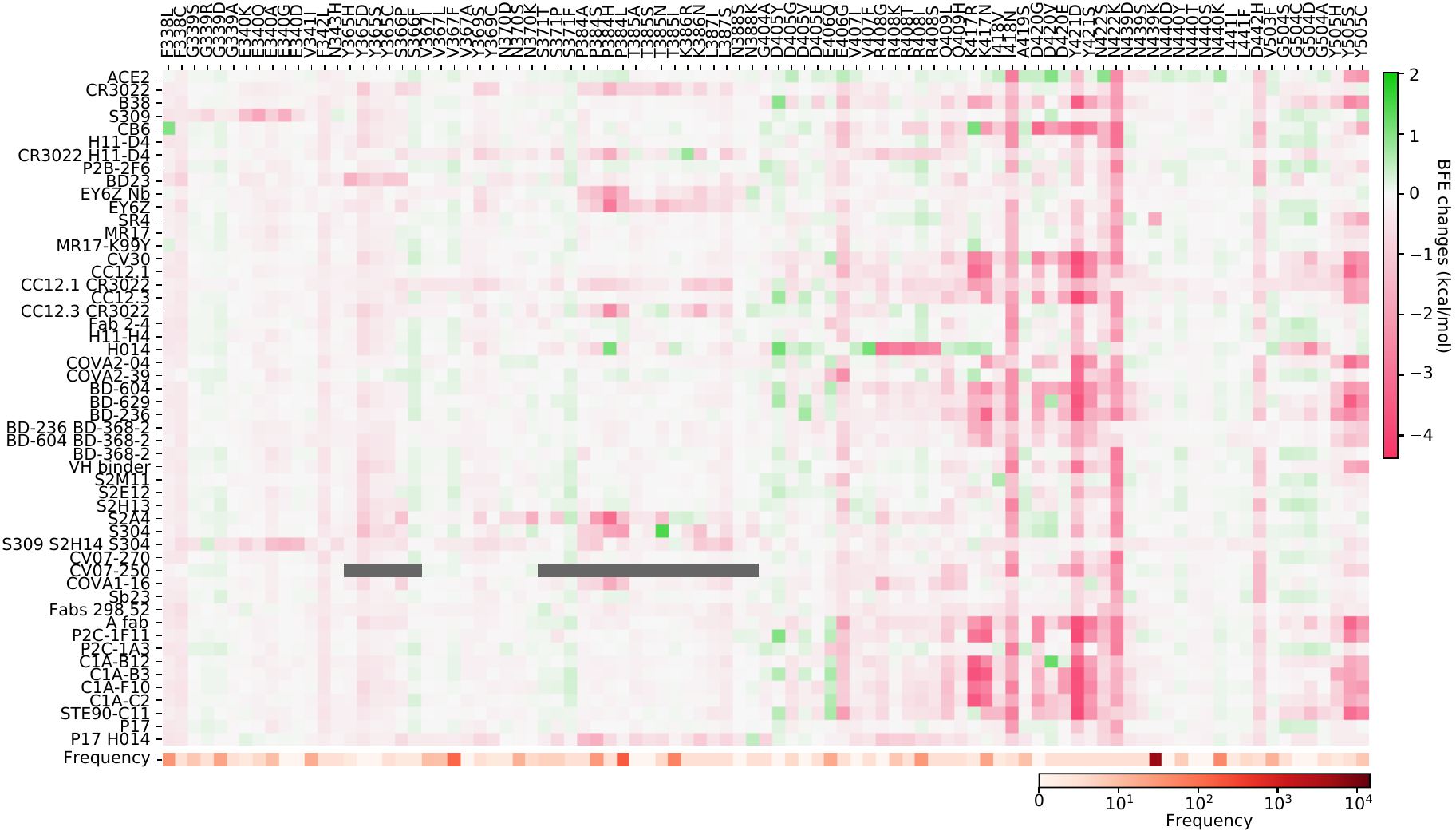}
\caption{Illustration of SARS-CoV-2 helix-residue mutation induced BFE changes for the complexes of S protein and 51 antibodies or ACE2. Positive changes strengthen the binding while negative changes weaken the binding. Mutation frequency is given  for each mutation.    Grey color indicates PDB structures does not include residues induced by those mutations.
}
\label{fig:heatmap_helix}
\end{figure}

In this section, we build a library of mutation-induced BFE changes for all mutations and all antibodies as well as ACE2. In principle, we could create a library of all possible mutations for all antibodies, as we did for ACE2 \cite{chen2020mutations}.  Here, we limit our effort to all existing mutations. Antibody 4A8 on the NTD has been discussed above. We consider antibodies on the RBD. 

Based on our earlier analysis, three types of SARS-CoV-2 S protein secondary structural residues have different mutation rates. Among them, the random coils are major components of the RDB and the NTD, as shown in Fig. \ref{fig:antibodies}. 
{ Most RBD mutations (287 of 462) occur on the residues whose secondary structures are coil, while   93 out of 462 mutations are on the helix, and  82 out of 462 mutations are on the sheet.} Therefore,  mutations on the RBD are split into three categories based on their locations in secondary structures helix, sheet, and coil. In Figure~\ref{fig:heatmap_helix}, we present the BFE changes for the complexes of the S protein and antibodies or ACE2 induced by  mutations on the helix residues of the S protein RBD. The frequency for each mutation is also presented. Most mutations on helix residues lead to negative BFE changes (pink squares), which weaken the bindings, whereas some mutations induce positive BFE changes (green squares). {It is noted that most mutations lead to the strengthening of the S protein and ACE2 binding, which is consistent   with the natural selection rule.  Mutations N406G, I418N, N422K, D442H, Y505S, and Y505C give rise to a strong weakening effect   on most antibodies.} The N439K mutation having the highest frequency, shows {a positive} BFE change on ACE2 {but negative changes on most} antibodies.  {Mutation D405Y appears to strengthen most antibodies.}

\begin{figure}[ht!]
\centering
\includegraphics[width=1\textwidth]{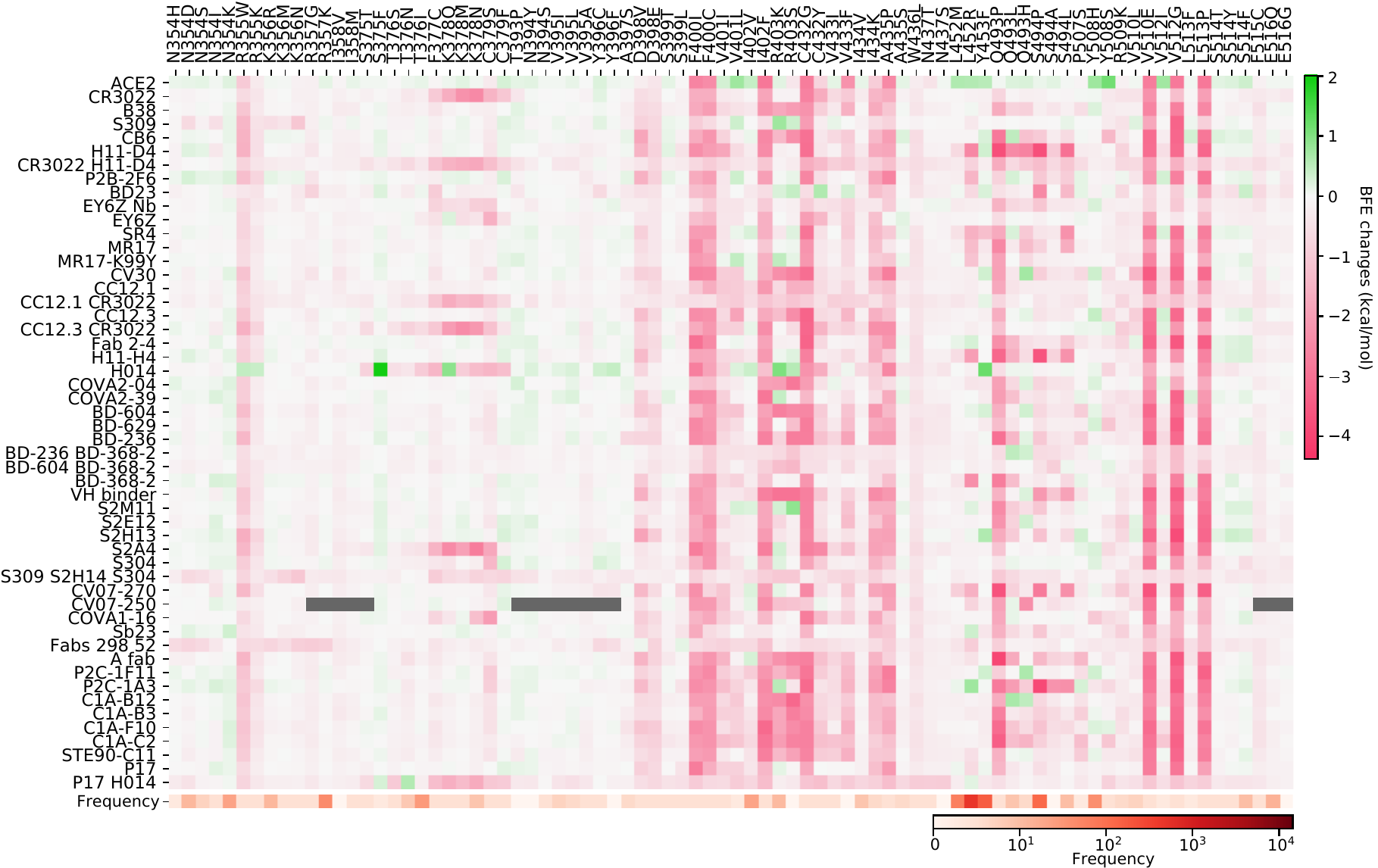}
\caption{Illustration of SARS-CoV-2 sheet-residue mutation induced BFE changes for the complexes of S protein and 51 antibodies or ACE2. Positive changes strengthen the binding while negative changes weaken the binding. Mutation frequency is presented for each mutation.  Grey color indicates PDB structures does not include residues induced by those mutations.
}
\label{fig:heatmap_sheet}
\end{figure}

In Figure~\ref{fig:heatmap_sheet}, we present the BFE changes for the S protein and antibody (ACE2) complexes following sheet residue mutations of the S protein RBD. {Like the last case, most mutations lead to positive BFE changes for ACE2, indicating  infectivity strengthening. There are many disruptive mutations, such as R355W, F401I, F401C, I402F, C432G, I434K, A435P, O493P, V510E, V512G, and L513P, that will weaken most antibodies and S protein complexes. 
On the other hand, most mutations strengthen certain antibodies but weaken other ones, which allows the effectiveness of antibody cocktails for a better protection.
The binding of antibody H014 and the S protein is strengthened by many mutations, particularly S375F, K378O, R403K, and Y453F. Among them,  Y453F is an infectivity-strengthening mutation with a relatively high frequency.
}


\begin{figure}[ht!]
	\centering
	\includegraphics[width=1\textwidth]{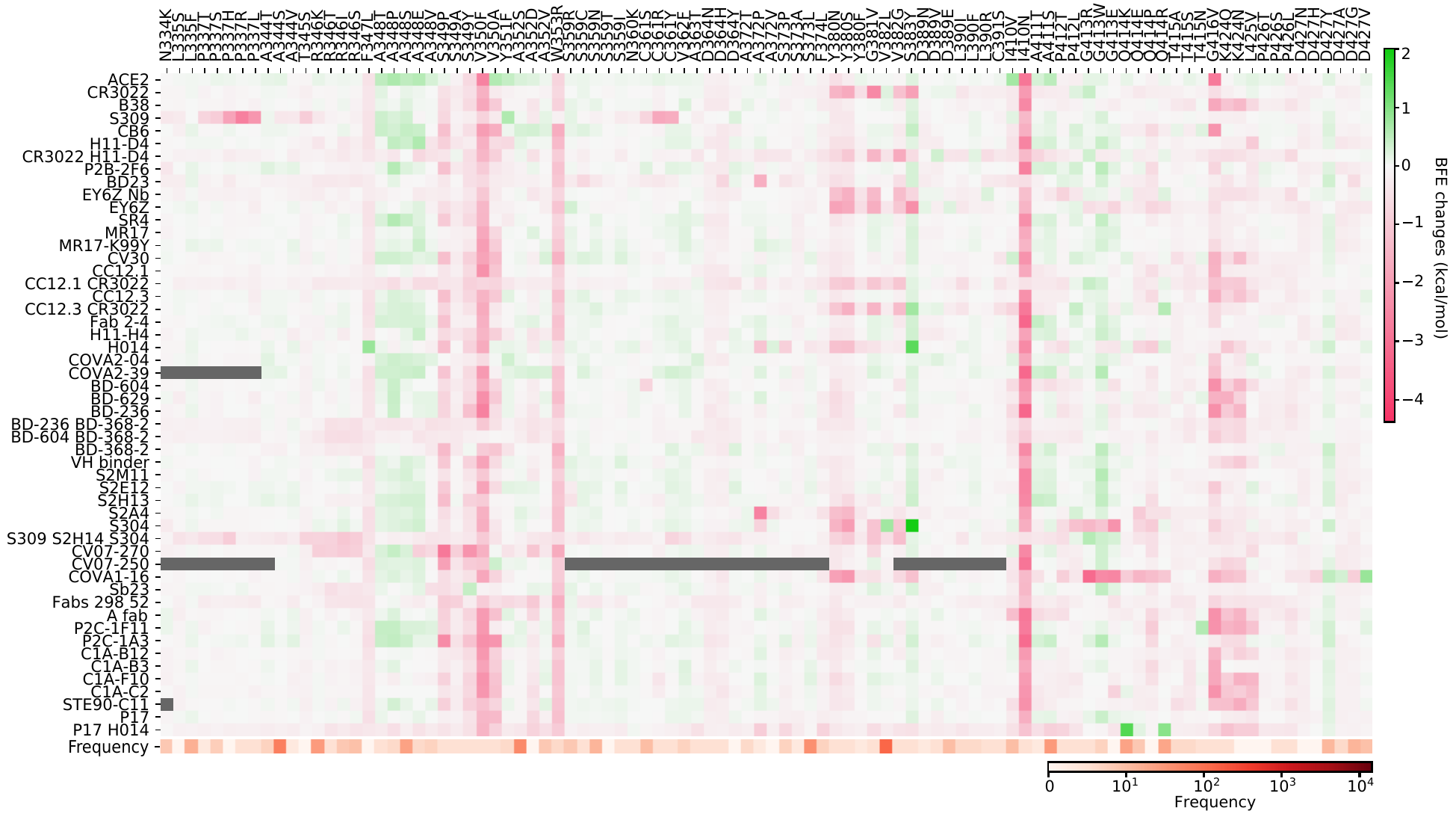}
	\caption{Illustration of SARS-CoV-2 coil-residue mutation induced BFE changes for the complexes of S protein and 51 antibodies or ACE2. Positive changes strengthen the binding while negative changes weaken the binding. Mutation frequency is presented for each mutation.  Grey color indicates PDB structures does not include residues induced by those mutations.}
	\label{fig:heatmap_coil_1}
\end{figure}

\begin{figure}[ht!]
\centering
\includegraphics[width=1\textwidth]{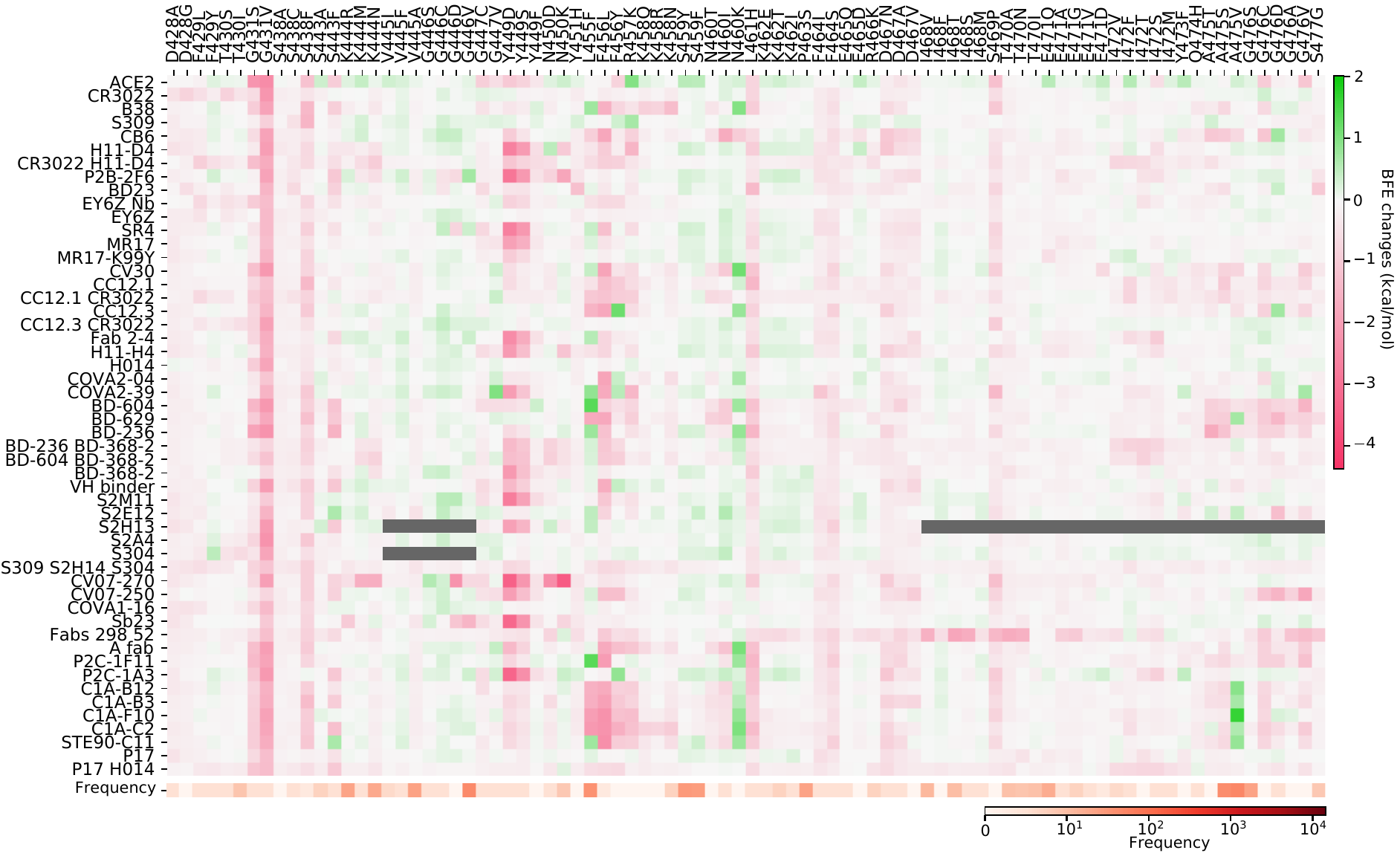}
\caption{Illustration of SARS-CoV-2 coil-residue mutation induced BFE changes for the complexes of S protein and 51 antibodies or ACE2 (continued from Fig. \ref{fig:heatmap_coil_1}). Positive changes strengthen the binding while negative changes weaken the binding. Mutation frequency is presented for each mutation.  Grey color indicates PDB structures does not include residues induced by those mutations.
}
\label{fig:heatmap_coil_2}
\end{figure}

\begin{figure}[ht!]
\centering
\includegraphics[width=1\textwidth]{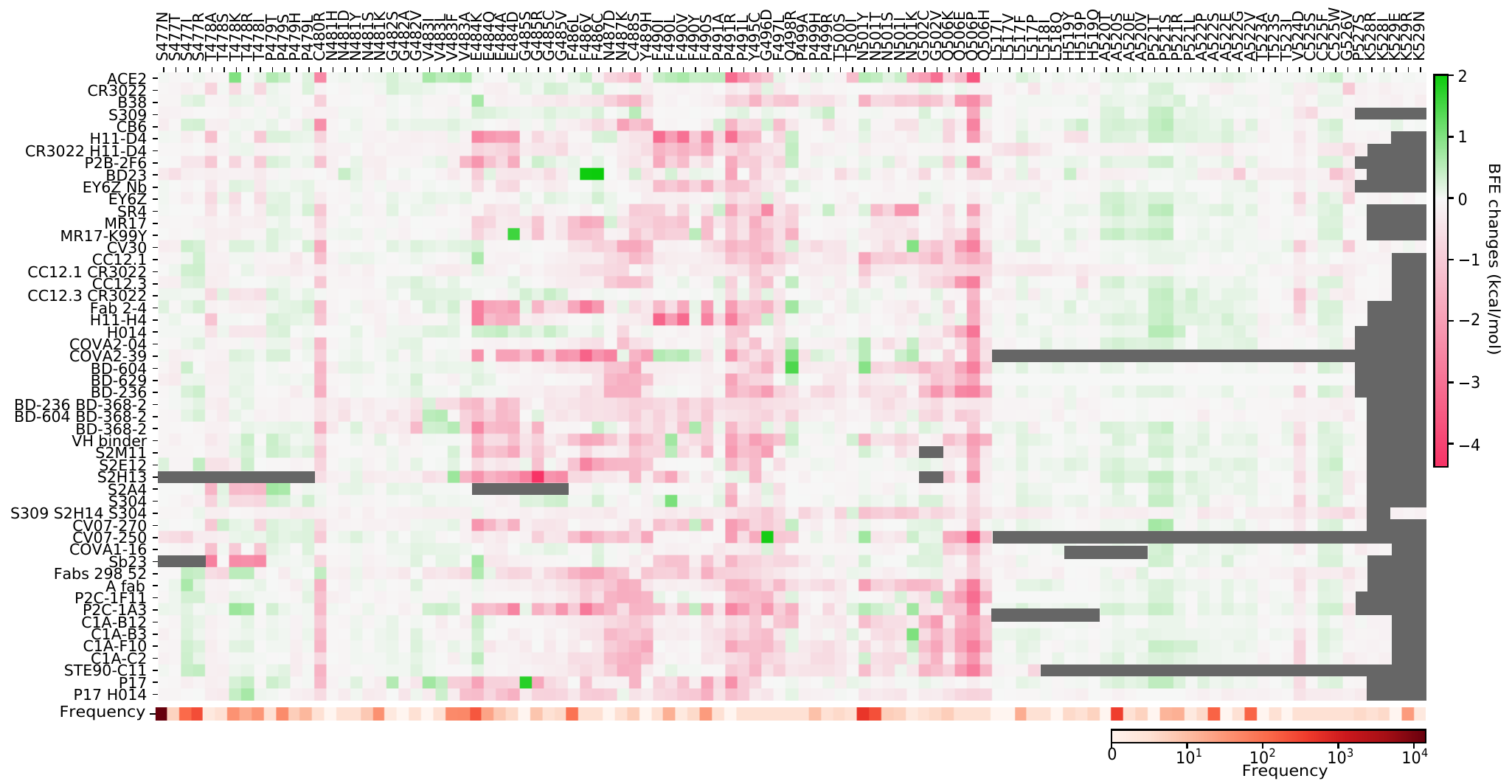}
\caption{Illustration of SARS-CoV-2 coil-residue mutation induced BFE changes for the complexes of S protein and 51 antibodies or ACE2 (continued from Fig. \ref{fig:heatmap_coil_2}). Positive changes strengthen the binding while negative changes weaken the binding. Mutation frequency is presented for each mutation.  Grey color indicates PDB structures does not include residues induced by those mutations.
}
\label{fig:heatmap_coil_3}
\end{figure}
Figures~\ref{fig:heatmap_coil_1}, \ref{fig:heatmap_coil_2}, and \ref{fig:heatmap_coil_3} 
present the BFE changes for the S protein and antibody (ACE2) complexes following coil residue mutations of the S protein RBD. Overall, most mutations on coil residues lead to {mild} negative BFE changes. 
 {However, mutations V350F, W353R, I401N, G416V, G431V, Y449D, Y449S, C480R, P491R, P491L, Y495C, and O506P will weaken most antibody bindings to the S protein. Some residues, like A348, N460, and P521, can produce many binding-strengthening mutations for most antibodies and ACE2.} For the high-frequency mutation S447N in Figure 
\ref{fig:heatmap_coil_3}, the BFE changes are mild on ACE2 and antibodies. 
 {Additionally,   the N501Y mutation, one of the typical mutations in the UK B.1.1.7 variants,  strengthens the infectivity but induces a mixed reactions to antibodies as shown in Figure \ref{fig:heatmap_coil_3} . }

\subsection{Statistical analysis of mutation impacts on COVID-19 antibodies}

\begin{figure}[ht!]
\centering
\includegraphics[width=0.8\textwidth]{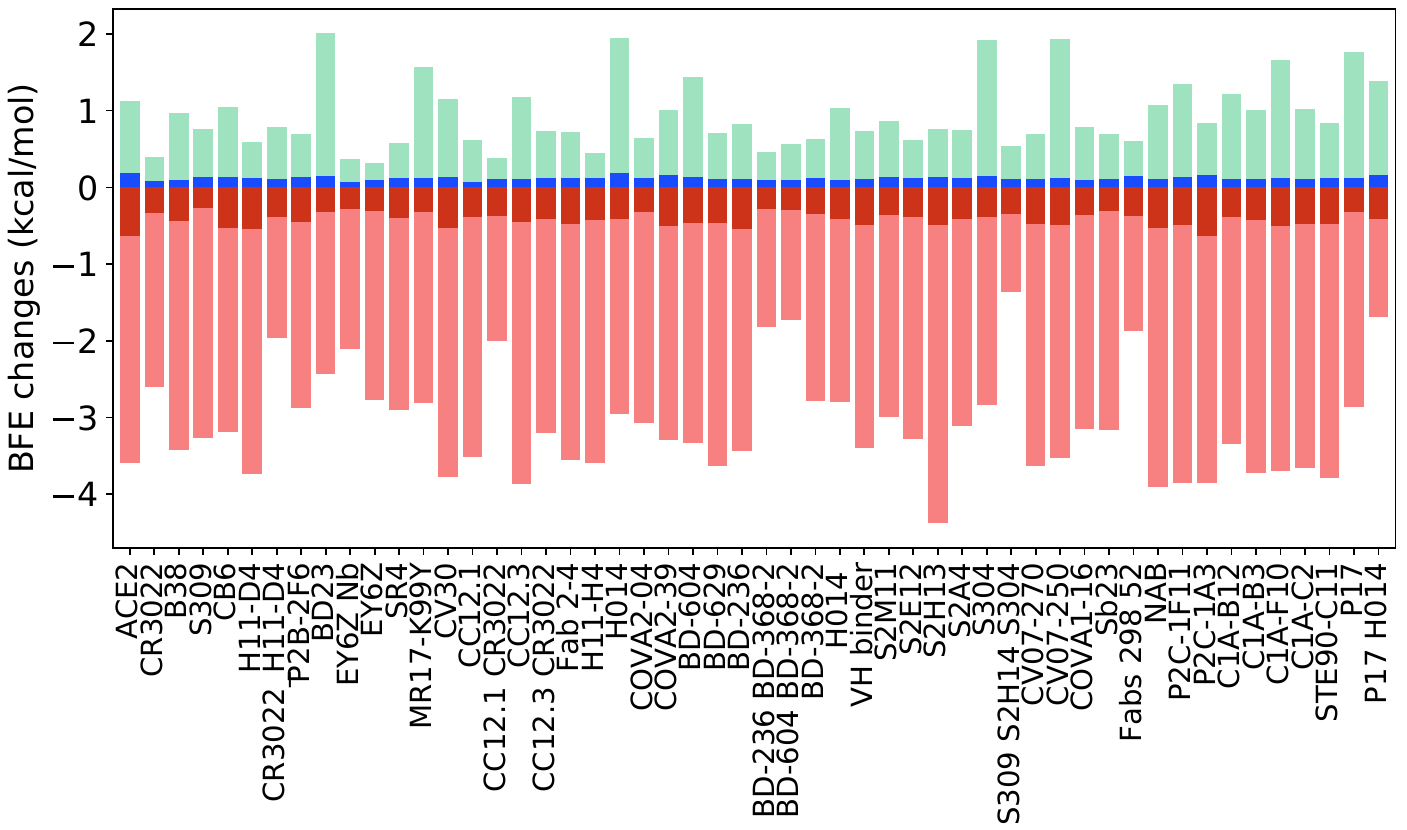}
\caption{Illustration of SARS-CoV-2 mutation-induced maximal and minimal BFE changes in cyan and pink for the complexes of S protein and 51 antibodies or ACE2, and average of positive and negative BFE changes in blue and red. Here, the maximal change strengthens the binding while the minimal change weakens the binding for each complex.
}
\label{fig:Barplot}
\end{figure}

{First, we perform a statistical analysis of all mutation-induced BFE changes studied in the last section. Most mutations induce binding-weakening BFE changes. The total rate of negative BFE changes is 71\% (i.e., 16661   out of 23512),  for coil residues, 67\% BFE changes are negative,  while for helix or sheet residues, 72\% and 80\% BFE changes are negative, respective. However, for ACE2, 300 out of 462 mutations (i.e., 65\%) on the REB produce positive or binding-strengthening  BFE changes, showing the effect of the natural selection of mutations. In contrast,  at most 200 of 462 mutations on the RBD give rise to negative BFE changes for antibodies. More specifically,  11 antibodies have less  than   100 positive BFE changes while 41 antibodies have less   than  200 positive BFE changes. Interestingly, in our prediction, 4 out of the 43 single antibodies have less  than   100 positive BFE changes, while 7 out of the 9 antibody cocktails have less   than  100 positive BFE changes. Although antibody cocktails have mild negative BFE changes, it turns out that they have high affinities to S protein and the BFE changes are mild for positive ones as well.}

Figure~\ref{fig:Barplot} indicates the BFE changes extreme values (maximal in cyan and minimal in pink) and average values (positive in blue and negative in red) of the complexes of S protein ACE2 or antibodies following mutations. 
{ The maximal  BFE changes of the helix, sheet, and coil residues  are 1.44 kcal/mol, 1.94 kcal/mol, and 1.00 kcal/mol, respectively,   while the minimal BFE changes are -3.87 kcal/mol, -3.9 kcal/mol, and -4.38 kcal/mol, respectively. The disparity in their maximal and minimal values indicates that relatively optimal nature of the S protein and antibody binding complexes. It means that the human immune system has the ability to produce optimized antibodies for a given antigen. However, antibodies, once generated, are prone to the infection of new mutants.  The disparity    shown in Figure~\ref{fig:Barplot}  also means that the SARS-CoV-2 was at an advanced stage of evolution with respect to human infection. There is not much room for SARS-CoV-2 to improve its infectivity by single-site mutations.  
 }

{Many antibody cocktails, such as CR3022/H11-D4, CC12.1/CR3022, BD-236/BD368-2, BD604/BD368-2, S309/S2H14/S304, and Fabs 298/52, are relatively less  sensitive to the current S protein mutations.} However, some other antibodies, such as H11-D4, CV30, CC12.3, and S2H13, can be dramatically affected by SARS-CoV-2 mutations. {Importantly, ACE2 is also impacted by mutations and has the largest positive BFE change    on average.}

\section{Mutation impacts on COVID-19 vaccines}

 The increasing number of infection cases and deaths, the global spread situation, and the lack of prophylactics and therapeutics give rise to the urgent demand for the prevention of COVID-19. Vaccination is the most effective and economical means to control pandemics \cite{zhang2020progress}. Currently, { 248} vaccines are in various clinical trial stages, as reported in an online COVID-19 Treatment And Vaccine Tracker (\url{https://covid-19tracker.milkeninstitute.org/#vaccines_intro}). Broadly speaking, there are four types of coronavirus vaccines in progress: virus vaccines, viral-vector vaccines, nucleic-acid vaccines, and protein-based vaccines, as shown in  \autoref{fig:vaccine}.  The first type of vaccine is the virus vaccine, which injects weakened or inactivate viruses into the human body. A virus is conventionally weakened by altering its genetic code to reduce its virulence and elicit a stronger immune response. A biotechnology company Codagenix is currently working on a ``codon optimization'' technology to weaken viruses, and it has weakened virus vaccine is in progress \cite{chen2020sars}. Unlike a weakened virus, the inactivated virus cannot replicate in the host cell. A virus is inactivated by heating or using chemicals, which induces neutralizing antibody titers and has been proven to have its safety \cite{lin2007safety}. At this stage, both Sinopharm, which works with the Beijing Institute of Biological Products and Wuhan Institute of Biological Products, and Sinovac, which works with Institute Butantan and Bio Farma is developing inactive SARS-CoV-2 vaccines that are in Phase III clinical trials.

The second type of vaccine is the viral-vector vaccine, which is genetically engineered so that it can produce coronavirus surface proteins in the human body without causing diseases. There are two subtypes of viral-vector vaccines: the non-replicating viral vector and the replicating viral vector. { On February 25, 2021, the World Health Organization (WHO) granted an emergency use listing (EUL) for the vaccine developed by AstraZeneca and the University of Oxford, which is a non-replicating viral vector vaccine. Moreover, there are 3 non-replicating viral vector vaccines in Phase III trials as well.} It works by taking a chimpanzee virus and coating it with the S proteins of SARS-CoV-2. The chimp virus causes a harmless infection in humans, but the spike proteins will activate the immune system to recognize signs of a future SARS-CoV-2 invasion. Notably, booster shots can be needed to keep long-lasting immunity. Furthermore, at this stage, only one replicating viral-vector vaccine is in Phase II. { The University of Hong Kong, in cooperating with the Xiamen University and Wantai Biological Pharmacy, is developing such a replicating viral vaccine, which tends to be safe and provoke a strong immune response.}

The third type of vaccines is nucleic acid vaccines, including two subtypes: DNA-based vaccines and RNA-based vaccines. At least { 40} teams are currently working on nucleic-acid vaccines since they are safe and easy to develop. The DNA-based vaccine works by inserting genetically engineered blueprints of the viral gene into small DNA molecules such as plasmids for injection. Moreover, the electroporation technique is employed to create pores in membranes to increase DNA uptake into cells. The injected DNA will produce mRNA by transcription with the help of the nucleus in human cells. Such an mRNA will translate viral proteins (mostly spike proteins), which are dutifully produced by cells in response to the genes, alarm the immune system, and should produce immunity. Currently, { there is one DNA-based vaccine in Phase III.} Similar to DNA-based vaccine, the RNA-based vaccine provides immunity through the introduction of RNA, which is encased in a lipid coat to ensure its entering into cells.  { Two RNA-based vaccines are granted authorization for emergency use in many countries. One is  designed by Biontec, which cooperates with Pfizer, and the other one is from the Moderna.  }

The fourth type of vaccine is the protein-based vaccine, which aims to inject viral proteins directly to human bodies into trigger immune readiness. Protein subunits vaccine is one of the subtypes of the protein-based vaccine. More than { 80} teams are working on vaccines with viral protein subunits, such as spike proteins and membrane (M) proteins. Another subtype of the protein-based vaccine is the virus-like particle (VLP) vaccine. The VLP vaccines closely resemble viruses. However, they are not infectious since they do not contain viral genetic material. The non-replicating propriety provides a safer alternative to weakened virus vaccines, the HPV vaccine or newer flu vaccines are VLP vaccines. Currently, { 22} teams are working on the VLP vaccines for the future prevention of COVID-19.

\subsection{Secondary structures of antigenic determinants}

\begin{figure}[H]
\centering
\includegraphics[width=0.8\textwidth]{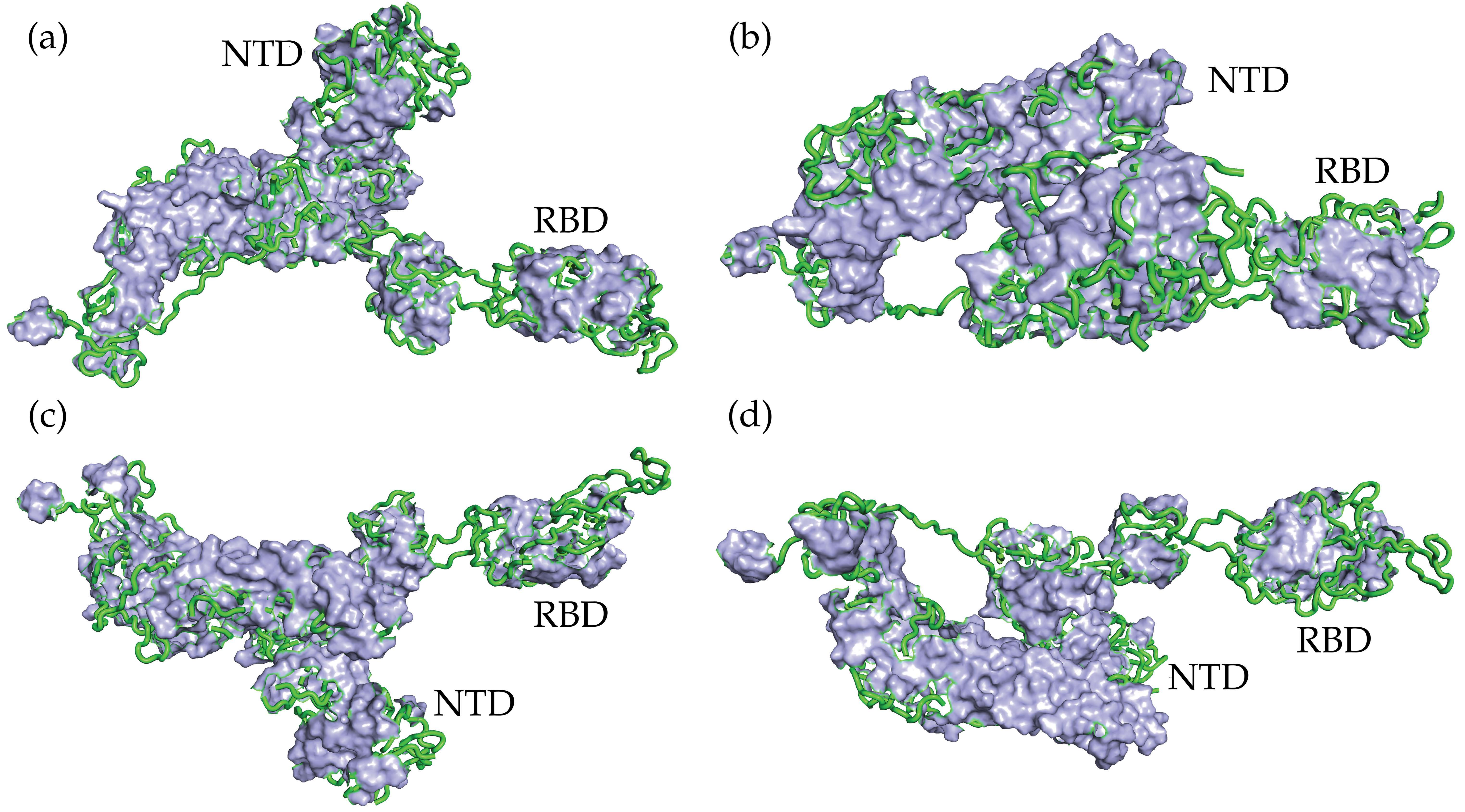}
\caption{The 3D rotational structure of SARS-CoV-2 S protein. The random coils of S protein is drawn with green strings and the other secondary structure is described with the purple surface. (a) 3D structure of S protein. (b) 3D structure of S protein that rotate 90 degree based on (a). (c) 3D structure of S protein that rotate 180 degree based on (a). (d) 3D structure of S protein that rotate 270 degree based on (a). }
\label{fig:surface_coils}
\end{figure}

Since the structural basis of antibody CDRs, or paratope,  is random coils, we hypothesize that CDRs favor antigenic random coils as complementary epitopes, i.e., antigenic determinants \cite{li2013bioinformatic,kringelum2013structural}.  \autoref{fig:surface_coils} depicts the 3D structure of S protein, where the random coils are drawn with green strings, and the other secondary structure is described with the purple surface. It shows that the RBD and the NTD mostly consist of random coils. The RBD is the antigenic determinant of { 43} structurally-known SARS-CoV-2 antibodies; meanwhile, the NTD is the binding domain of  {{ antibodies 4A8 and FC05  and antibody 2G12 also binds to the S2 domain with random coils}}, which confirms our hypothesis. {\color {red} More detailed analysis considered the random coil percentages of antibodies' paratope and are summarized in Table S1 of the Supporting material. It reveals antibodies predominantly contact residues in random coils of S protein. In the most case, the percentages are over 90 \%.}   

\begin{figure}[H]
    \centering
    \includegraphics[width=1\textwidth]{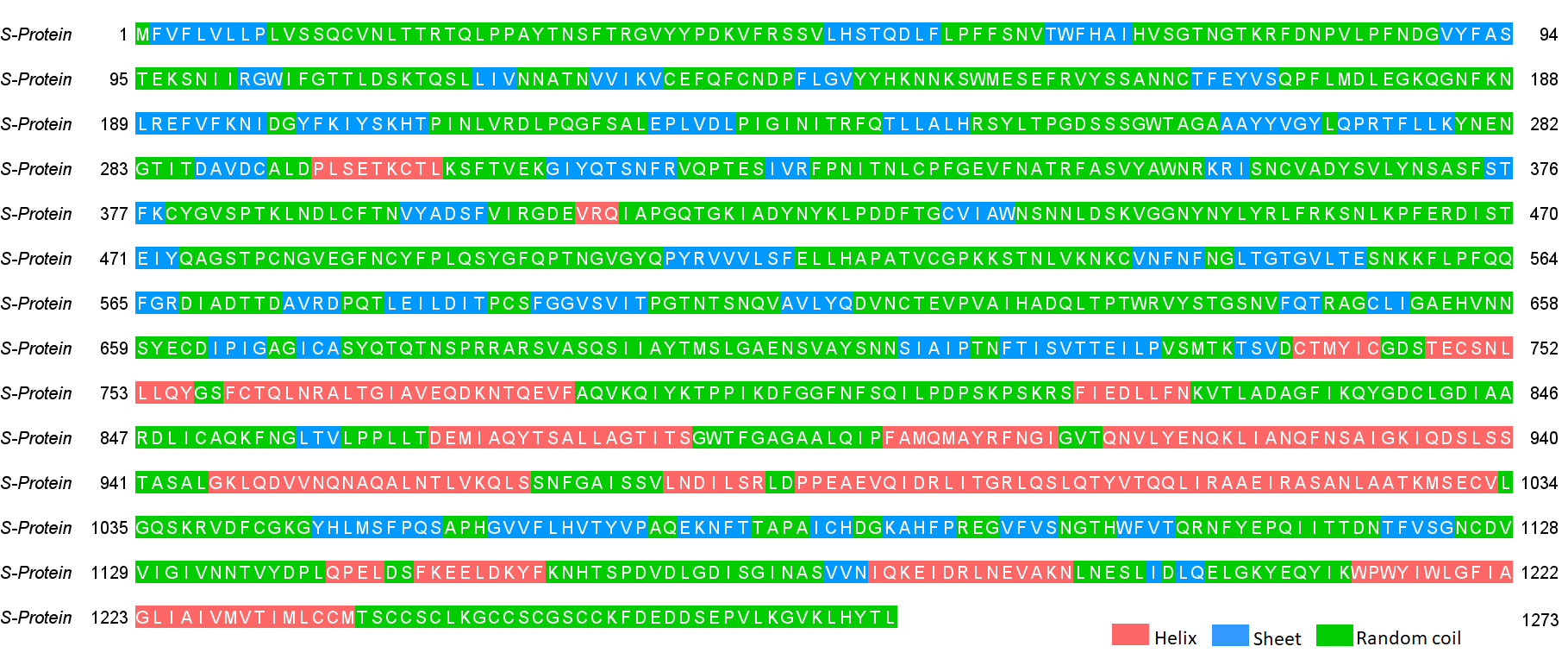}
    \caption{The secondary structure of S protein. The red, green, and blue colors represent helix, sheet, and random coils of S protein. }
    \label{fig:second structure}
\end{figure}

\autoref{fig:second structure} marks the secondary structure of the S protein. The red, blue, and green colors represent helix, sheet, and random coils of S protein. It can be seen that the S protein mostly consists of random coils, which means there are many other potential antigenic epitopes on the S protein for antibody CDRs. We believe that the emphasis   on direct binding competition with ACE2 in the past \cite{pinto2020structural, tian2020potent,zhou2020structural} has led to the neglecting of many important antibodies that do not bind to the RBD. Therefore, we suggest that researchers pay more attention to antibodies that do not bind to the RBD.

\subsection{Statistical estimation of mutation impacts on COVID-19 vaccines}

Vaccine efficacy is an essential issue for the control of the COVID-19 pandemic. The S protein is one of the most popular surface proteins for vaccine development. However, mutations accumulated on the S protein of SARS-CoV-2, which may reduce the vaccine efficacy. As we found in \autoref{sec:S protein}, mutations are more likely to happen on the random coils of S protein, which may have a devastating effect on vaccines in the development. 
 
 As shown in \autoref{fig:Barplot}, mutations could considerably weaken the binding between the S protein and antibodies and thus pose a direct threat to reduce the efficacy of vaccines. However, there are a few obstacles in determining the exact impacts of mutations   on COVID-19 vaccines. Firstly, the four types of vaccine platforms can produce very different virus peptides, resulting in different immune responses, as well as antibodies. Secondly, even for a given vaccine platform, the different peptides may be produced due to different immune responses caused by gender difference, age difference, race difference, etc. Therefore, in this work, we proposed to understand the impact of SARS-CoV-2 mutations on COVID-19 vaccines by the statistical analysis. By evaluating the binding affinity changes induced by 51 existing SARS-CoV-2 antibodies, as shown in \autoref{fig:heatmap_helix} to \autoref{fig:heatmap_coil_3}, 
{ we can identify vaccine escape mutants. Table \ref{tab:EscapeMutants} list a collection of the most disruptive mutations. However, this list is not complete. There are many other vaccine escape mutations as shown in  \autoref{fig:heatmap_helix} to \autoref{fig:heatmap_coil_3}.
For example, the infectivity-strengthening South Africa mutant E484K can cause dramatically disruptive effects on many antibodies such as H11-D4, Fab 2-4, H11-H4, COVA2-39, BD368-2, etc. but it also enhances the binding of other antibodies, such as B38, CV30, CC21.1, Sb23, Fabs 298 52, etc. The infectivity-strengthening   mutation N501Y in UK B.1.1.7 variants has a disruptive effect only on a few known antibodies,  including B38, CC12.3, S2M11, NAB, S309 S2H12 S304, C1A-B12, STE90-C11, etc.    
}
\begin{table}[ht!]
    \centering
    \setlength\tabcolsep{11pt} 
	\captionsetup{margin=0.1cm}
	\caption{Vaccine escape mutants}
    \label{tab:EscapeMutants}
    \begin{tabular}{c|c}
    \toprule
   Location &  Mutants  \\ 
    \midrule
    Helix        &  E406G, I418N, Y421D, N422K, D442H, Y505S \\ 
    Sheet        &  R355W, F400I, F400C, I402F, C432G, I434K, A435P, Q493P, V510E, V512G, L513P   \\ 
    Coils        &  V350F, W353R, I410N, G416V, G431V, Y449D, Y449S, L461H, S469P, C480R, P491R,  \\ & P491L, Y495C, Q506P   \\ 
    \bottomrule
    \end{tabular}
\end{table}


 In a nutshell, by setting up a SARS-CoV-2 antibody library with the statistical analysis based on the mutation-induced binding free energies changes, we can estimate the impacts of SARS-CoV-2 mutations on COVID-19 vaccines, which will provide a way to infer how a specific mutation will pose a threat to vaccines. This approach works better when more antibody structures become available. 

Another important factor in prioritization is mutation frequency.  Figures~\ref{fig:heatmap_helix}, \ref{fig:heatmap_sheet},   \ref{fig:heatmap_coil_1},   \ref{fig:heatmap_coil_2},
and  \ref{fig:heatmap_coil_3} have provided frequency information from our SNP calling. Once a mutation is identified as a potential threat, it can be incorporated into the next generation of vaccines in a cocktail approach. In principle, all four types of vaccine platforms allow the accommodation of new viral strains. 

\section{Validation}

{
Although the details of the methods used in this work are presented in the Supporting  material, we provide a validation of our deep learning prediction model, TopNetTree \cite{wang2020topology}, which is crucial to the credibility of this work. 
Specifically, we demonstrate the prediction performance of S protein mutation induced BEF changes on CTC-445.2 compared to the experimental deep mutations enrichment data \cite{linsky2020novo}. More detailed descriptions of methods and datasets are illustrated in the Supporting material.}

\begin{figure}[H]
	\centering
	\includegraphics[width=1\textwidth]{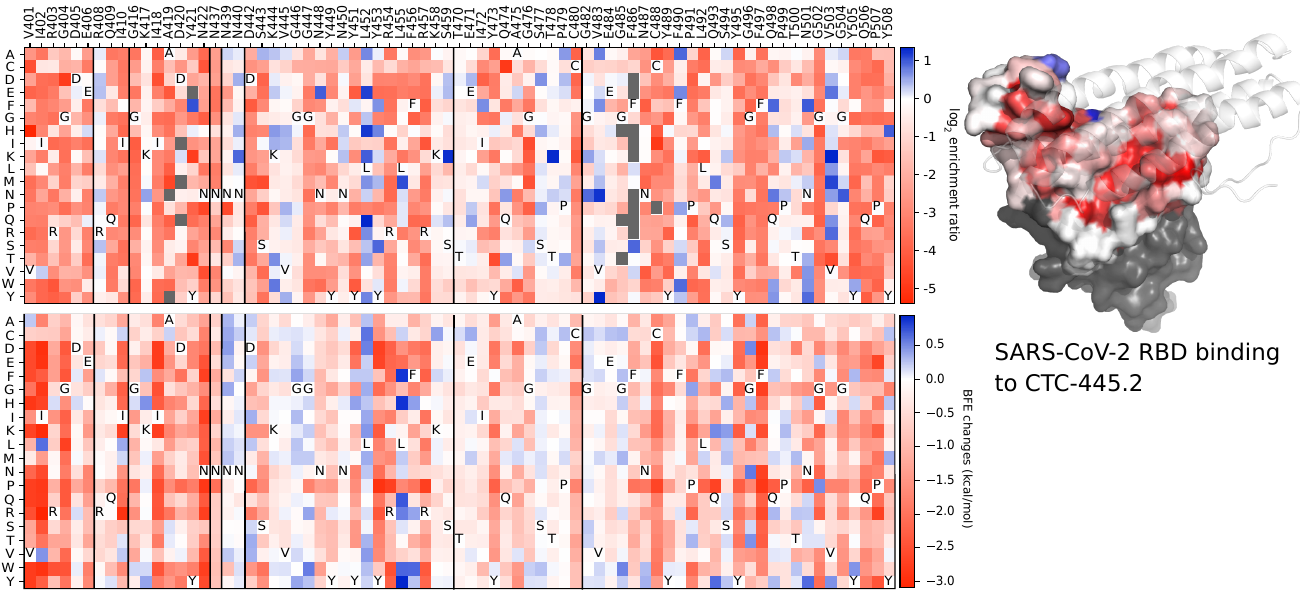}
	\caption{{A comparison between experimental deep mutation enrichment data and TopNetTree predictions for SARS-CoV-2 S protein RBD and CTC-445.2  complex (7KL9 \cite{linsky2020novo}). \textbf{Top left}: deep mutational scanning heatmap showing the average effect on the enrichment for single site mutants of the RBD when assayed by yeast display for binding to   CTC-445.2 \cite{linsky2020novo}. \textbf{Top right}: the RBD colored by average enrichment at each residue position bound to CTC-445.2. \textbf{Bottom}: machine learning predicted BFE changes for CTC-445.2 and S protein complex induced by single site mutations on the RBD.}}
	\label{fig:7KL9_RBD_combine}
\end{figure}
{Figure \ref{fig:7KL9_RBD_combine} presents a comparison between experimental deep mutation enrichment data on the RBD and machine learning predicted RBD-mutation-induced BFE changes for the SARS-CoV-2 S protein and CTC-445.2 complex. In the heatmaps of Figure \ref{fig:7KL9_RBD_combine}, one can see that the predicted BFE changes have a very high correlation to the experimental enrichment ratio data. Both enrichment ratios and BFE changes describe the affinity strength of the protein-protein interaction induced by mutations. The high similarity between these heatmaps demonstrates the reliability of  our machine learning predictions of BFE changes following mutations on the S protein RBD.  
 }

\section{Conclusion}
 
Coronavirus disease 2019 (COVID-19) pandemic has gone out of control globally. There is no specific medicine and effective treatment for this viral infection at this point. Vaccination is widely anticipated to be the endgame for taming the viral rampant. Another promising treatment that is relatively easy to develop is antibody therapies. However, both vaccines and antibody therapies are prone to more than { 26,000} unique mutations recorded in the \href{https://users.math.msu.edu/users/weig/SARS-CoV-2_Mutation_Tracker.html}{Mutation Tracker}.

We present {the most comprehensive analysis and prediction} of mutation threats to vaccines and antibody therapies. First, we identify existing mutations on the severe acute respiratory syndrome coronavirus 2 (SARS-CoV-2) spike (S) protein, which is the main target for both vaccines and antibody therapies. 
We analyze the mechanism, frequency, and ratio of mutations along with the secondary structures of the S protein. Additionally, we build a library of { 55} antibodies with structures available from the Protein Data Bank (PDB) and analyze their two-dimensional (2D) and three-dimensional (3D) characteristics by employing computational biophysics. We further predict the mutation-induced binding free energy (BFE) changes of S protein and antibody complexes by a model called TopNetTree based on deep learning and algebraic topology. { The performance of our model has been extensively validated by its prediction of experimental deep mutation data.}   {
Our significant findings are the following.
First, we reveal that none of the known mutations is safe to all antibodies. On average, most mutations (i.e., 71\%) will weaken the binding between the S protein and antibodies, which implies that vaccines will also be compromised by existing mutations.    
Additionally, we identify  31 vaccine escape mutants that dramatically weaken the binding between the S protein and most known antibodies.  Moreover, we find that most RBD mutations (i.e., 64.9\%) will enhance the binding strength between the S protein and angiotensin-converting enzyme 2 (ACE2), which implies most existing mutations will strengthen the SARS-CoV-2 infectivity. This result is consistent with the natural selection of mutations and our earlier finding. \cite{chen2020mutations}
Finally, we discover that the maximal BFE change magnitudes of binding-strengthening mutations are much smaller than those of binding-weakening mutations for all antibodies, which shows current human antibodies were optimized with respect to the original S protein and are prone to the S protein mutations. Our findings indicate the pressing need to keep developing mutation-resistant vaccines and antibody drugs and to be ready for seasonal vaccinations.   
}   

\section*{Supporting material} 
 Supporting material is available for: 
S1 Methods;  
S2 Multiple sequence alignments of antibodies and pairwise identity scores; {
S3 Random coil percentages of antibody paratopes; and 
S4 Additional analysis of antibody-S protein complexes.}

\section*{Data availability}  
Detailed mutation information is available for download at 
\href{https://users.math.msu.edu/users/weig/SARS-CoV-2_Mutation_Tracker.html}{Mutation Tracker}.

\section*{Acknowledgments}
This work was supported in part by NIH grant  GM126189, NSF grants DMS-2052983,  DMS-1761320, and IIS-1900473,  NASA grant 80NSSC21M0023,  Michigan Economic Development Corporation,  George Mason University award PD45722,  Bristol-Myers Squibb 65109, and Pfizer.
The authors thank The IBM TJ Watson Research Center, The COVID-19 High Performance Computing Consortium, NVIDIA, and MSU HPCC for computational assistance. RW thanks Dr. Changchuan Yin for useful discussion.  

\section*{Competing Interests}
The authors declare no competing interests.


\begin{thebibliography}{10}

\bibitem{lu2020genomic}
Roujian Lu, Xiang Zhao, Juan Li, Peihua Niu, Bo~Yang, Honglong Wu, Wenling
  Wang, Hao Song, Baoying Huang, Na~Zhu, et~al.
\newblock Genomic characterisation and epidemiology of 2019 novel coronavirus:
  implications for virus origins and receptor binding.
\newblock {\em The Lancet}, 395(10224):565--574, 2020.

\bibitem{shin2020covid}
Matthew~D Shin, Sourabh Shukla, Young~Hun Chung, Veronique Beiss, Soo~Khim
  Chan, Oscar~A Ortega-Rivera, David~M Wirth, Angela Chen, Markus Sack,
  Jonathan~K Pokorski, et~al.
\newblock {COVID-19} vaccine development and a potential nanomaterial path
  forward.
\newblock {\em Nature Nanotechnology}, pages 1--10, 2020.

\bibitem{day2020covid}
Michael Day.
\newblock Covid-19: four fifths of cases are asymptomatic, {C}hina figures
  indicate, 2020.

\bibitem{long2020clinical}
Quan-Xin Long, Xiao-Jun Tang, Qiu-Lin Shi, Qin Li, Hai-Jun Deng, Jun Yuan,
  Jie-Li Hu, Wei Xu, Yong Zhang, Fa-Jin Lv, et~al.
\newblock Clinical and immunological assessment of asymptomatic {SARS-CoV-2}
  infections.
\newblock {\em Nature medicine}, 26(8):1200--1204, 2020.

\bibitem{kissler2020projecting}
Stephen~M Kissler, Christine Tedijanto, Edward Goldstein, Yonatan~H Grad, and
  Marc Lipsitch.
\newblock Projecting the transmission dynamics of {SARS-CoV-2} through the
  postpandemic period.
\newblock {\em Science}, 368(6493):860--868, 2020.

\bibitem{12yeartrip}
{12-year trip}.
\newblock
  \url{https://www.medicinenet.com/script/main/art.asp?articlekey=9877}.

\bibitem{bloch2020deployment}
Evan~M Bloch, Shmuel Shoham, Arturo Casadevall, Bruce~S Sachais, Beth Shaz,
  Jeffrey~L Winters, Camille van Buskirk, Brenda~J Grossman, Michael Joyner,
  Jeffrey~P Henderson, et~al.
\newblock Deployment of convalescent plasma for the prevention and treatment of
  {COVID-19}.
\newblock {\em The Journal of clinical investigation}, 130(6):2757--2765, 2020.

\bibitem{prompetchara2020immune}
Eakachai Prompetchara, Chutitorn Ketloy, and Tanapat Palaga.
\newblock Immune responses in {COVID-19} and potential vaccines: Lessons
  learned from {SARS} and {MERS} epidemic.
\newblock {\em Asian Pac J Allergy Immunol}, 38(1):1--9, 2020.

\bibitem{wu2020neutralizing}
Fan Wu, Aojie Wang, Mei Liu, Qimin Wang, Jun Chen, Shuai Xia, Yun Ling, Yuling
  Zhang, Jingna Xun, Lu~Lu, et~al.
\newblock Neutralizing antibody responses to {SARS-CoV-2} in a {COVID-19}
  recovered patient cohort and their implications.
\newblock {\em medRxiv}, 2020.

\bibitem{li2020orf6}
Jin-Yan Li, Ce-Heng Liao, Qiong Wang, Yong-Jun Tan, Rui Luo, Ye~Qiu, and
  Xing-Yi Ge.
\newblock The {ORF6, ORF8} and nucleocapsid proteins of {SARS-CoV-2} inhibit
  type {I} interferon signaling pathway.
\newblock {\em Virus research}, 286:198074, 2020.

\bibitem{tufan2020covid}
Abdurrahman Tufan, ASLIHAN~AVANO{\u{G}}LU G{\"U}LER, and Marco Matucci-Cerinic.
\newblock {COVID-19}, immune system response, hyperinflammation and repurposing
  antirheumatic drugs.
\newblock {\em Turkish Journal of Medical Sciences}, 50(SI-1):620--632, 2020.

\bibitem{liang2020highlight}
Yanwen Liang, Mong-Lien Wang, Chian-Shiu Chien, Aliaksandr~A Yarmishyn, Yi-Ping
  Yang, Wei-Yi Lai, Yung-Hung Luo, Yi-Tsung Lin, Yann-Jang Chen, Pei-Ching
  Chang, et~al.
\newblock Highlight of immune pathogenic response and hematopathologic effect
  in {SARS-CoV}, {MERS-CoV}, and {SARS-CoV-2} infection.
\newblock {\em Frontiers in Immunology}, 11:1022, 2020.

\bibitem{catanzaro2020immune}
Michele Catanzaro, Francesca Fagiani, Marco Racchi, Emanuela Corsini, Stefano
  Govoni, and Cristina Lanni.
\newblock Immune response in {COVID-19}: addressing a pharmacological challenge
  by targeting pathways triggered by {SARS-CoV-2}.
\newblock {\em Signal Transduction and Targeted Therapy}, 5(1):1--10, 2020.

\bibitem{chaplin2010overview}
David~D Chaplin.
\newblock Overview of the immune response.
\newblock {\em Journal of Allergy and Clinical Immunology}, 125(2):S3--S23,
  2010.

\bibitem{kumar2011pathogen}
Himanshu Kumar, Taro Kawai, and Shizuo Akira.
\newblock Pathogen recognition by the innate immune system.
\newblock {\em International reviews of immunology}, 30(1):16--34, 2011.

\bibitem{takeuchi2010pattern}
Osamu Takeuchi and Shizuo Akira.
\newblock Pattern recognition receptors and inflammation.
\newblock {\em Cell}, 140(6):805--820, 2010.

\bibitem{kumar2009pathogen}
Himanshu Kumar, Taro Kawai, and Shizuo Akira.
\newblock Pathogen recognition in the innate immune response.
\newblock {\em Biochemical Journal}, 420(1):1--16, 2009.

\bibitem{pancer2006evolution}
Zeev Pancer and Max~D Cooper.
\newblock The evolution of adaptive immunity.
\newblock {\em Annu. Rev. Immunol.}, 24:497--518, 2006.

\bibitem{hewitt2003mhc}
Eric~W Hewitt.
\newblock The {MHC} class {I} antigen presentation pathway: strategies for
  viral immune evasion.
\newblock {\em Immunology}, 110(2):163--169, 2003.

\bibitem{harty2000cd8+}
John~T Harty, Amy~R Tvinnereim, and Douglas~W White.
\newblock {CD8+ T} cell effector mechanisms in resistance to infection.
\newblock {\em Annual review of immunology}, 18(1):275--308, 2000.

\bibitem{ting2002genetic}
Jenny Pan-Yun Ting and John Trowsdale.
\newblock Genetic control of {MHC} class {II} expression.
\newblock {\em Cell}, 109(2):S21--S33, 2002.

\bibitem{alberts2018molecular}
Bruce Alberts, Alexander Johnson, Julian Lewis, David Morgan, Martin Raff,
  Peter~Walter Keith~Roberts, et~al.
\newblock Molecular biology of the cell.
\newblock 2018.

\bibitem{hu2020cytokine}
Biying Hu, Shaoying Huang, and Lianghong Yin.
\newblock The cytokine storm and {COVID}-19.
\newblock {\em Journal of medical virology}, 2020.

\bibitem{grewal1998cd40}
Iqbal~S Grewal and Richard~A Flavell.
\newblock {CD}40 and {CD}154 in cell-mediated immunity.
\newblock {\em Annual review of immunology}, 16(1):111--135, 1998.

\bibitem{crotty2004immunological}
Shane Crotty and Rafi Ahmed.
\newblock Immunological memory in humans.
\newblock In {\em Seminars in immunology}, volume~16, pages 197--203. Elsevier,
  2004.

\bibitem{putnam1979primary}
Frank~W Putnam, YS~Liu, and TL~Low.
\newblock Primary structure of a human {IgA1} immunoglobulin. {IV}.
  streptococcal {IgA1} protease, digestion, {Fab} and {Fc} fragments, and the
  complete amino acid sequence of the alpha 1 heavy chain.
\newblock {\em Journal of Biological Chemistry}, 254(8):2865--2874, 1979.

\bibitem{wang2007antibody}
Wei Wang, Satish Singh, David~L Zeng, Kevin King, and Sandeep Nema.
\newblock Antibody structure, instability, and formulation.
\newblock {\em Journal of pharmaceutical sciences}, 96(1):1--26, 2007.

\bibitem{hamers1993naturally}
CTSG Hamers-Casterman, T~Atarhouch, S~Muyldermans, G~Robinson, C~Hammers,
  E~Bajyana Songa, N~Bendahman, and R~Hammers.
\newblock Naturally occurring antibodies devoid of light chains.
\newblock {\em Nature}, 363(6428):446--448, 1993.

\bibitem{van1999comparison}
RHJ Van~der Linden, LGJ Frenken, B~De~Geus, MM~Harmsen, RC~Ruuls, W~Stok,
  L~De~Ron, S~Wilson, P~Davis, and CT~Verrips.
\newblock Comparison of physical chemical properties of llama {VHH} antibody
  fragments and mouse monoclonal antibodies.
\newblock {\em Biochimica et Biophysica Acta (BBA)-Protein Structure and
  Molecular Enzymology}, 1431(1):37--46, 1999.

\bibitem{forsman2008llama}
Anna Forsman, Els Beirnaert, Marl{\'e}n~MI Aasa-Chapman, Bart Hoorelbeke,
  Karolin Hijazi, Willie Koh, Vanessa Tack, Agnieszka Szynol, Charles Kelly,
  Aine McKnight, et~al.
\newblock Llama antibody fragments with cross-subtype human immunodeficiency
  virus type 1 ({HIV}-1)-neutralizing properties and high affinity for {HIV}-1
  gp120.
\newblock {\em Journal of virology}, 82(24):12069--12081, 2008.

\bibitem{hoffmann2020sars}
Markus Hoffmann, Hannah Kleine-Weber, Simon Schroeder, Nadine Kr{\"u}ger, Tanja
  Herrler, Sandra Erichsen, Tobias~S Schiergens, Georg Herrler, Nai-Huei Wu,
  Andreas Nitsche, et~al.
\newblock {SARS-CoV-2} cell entry depends on {ACE2} and {TMPRSS2} and is
  blocked by a clinically proven protease inhibitor.
\newblock {\em Cell}, 2020.

\bibitem{cao2020covid}
Xuetao Cao.
\newblock {COVID-19}: immunopathology and its implications for therapy.
\newblock {\em Nature reviews immunology}, 20(5):269--270, 2020.

\bibitem{chen2020convalescent}
Long Chen, Jing Xiong, Lei Bao, and Yuan Shi.
\newblock Convalescent plasma as a potential therapy for {COVID-19}.
\newblock {\em The Lancet Infectious Diseases}, 20(4):398--400, 2020.

\bibitem{shen2020treatment}
Chenguang Shen, Zhaoqin Wang, Fang Zhao, Yang Yang, Jinxiu Li, Jing Yuan,
  Fuxiang Wang, Delin Li, Minghui Yang, Li~Xing, et~al.
\newblock Treatment of 5 critically ill patients with {COVID-19} with
  convalescent plasma.
\newblock {\em Jama}, 323(16):1582--1589, 2020.

\bibitem{zhang2020progress}
Jinyong Zhang, Hao Zeng, Jiang Gu, Haibo Li, Lixin Zheng, and Quanming Zou.
\newblock Progress and prospects on vaccine development against {SARS-CoV-2}.
\newblock {\em Vaccines}, 8(2):153, 2020.

\bibitem{callaway2020race}
Ewen Callaway.
\newblock The race for coronavirus vaccines: a graphical guide.
\newblock {\em Nature}, 580(7805):576, 2020.

\bibitem{wu2020new}
Fan Wu, Su~Zhao, Bin Yu, Yan-Mei Chen, Wen Wang, Zhi-Gang Song, Yi~Hu, Zhao-Wu
  Tao, Jun-Hua Tian, Yuan-Yuan Pei, et~al.
\newblock A new coronavirus associated with human respiratory disease in
  {China}.
\newblock {\em Nature}, 579(7798):265--269, 2020.

\bibitem{sevajol2014insights}
Marion Sevajol, Lorenzo Subissi, Etienne Decroly, Bruno Canard, and Isabelle
  Imbert.
\newblock Insights into {RNA} synthesis, capping, and proofreading mechanisms
  of {SARS}-coronavirus.
\newblock {\em Virus research}, 194:90--99, 2014.

\bibitem{ferron2018structural}
Fran{\c{c}}ois Ferron, Lorenzo Subissi, Ana Theresa~Silveira De~Morais, Nhung
  Thi~Tuyet Le, Marion Sevajol, Laure Gluais, Etienne Decroly, Clemens
  Vonrhein, G{\'e}rard Bricogne, Bruno Canard, et~al.
\newblock Structural and molecular basis of mismatch correction and ribavirin
  excision from coronavirus {RNA}.
\newblock {\em Proceedings of the National Academy of Sciences},
  115(2):E162--E171, 2018.

\bibitem{wang2020decoding}
Rui Wang, Yuta Hozumi, Changchuan Yin, and Guo-Wei Wei.
\newblock Decoding {SARS-CoV-2} transmission, evolution and ramification on
  {COVID-19} diagnosis, vaccine, and medicine.
\newblock {\em arXiv preprint arXiv:2004.14114}, 2020.

\bibitem{wang2020decoding0}
Rui Wang, Yuta Hozumi, Changchuan Yin, and Guo-Wei Wei.
\newblock Decoding {SARS-CoV-2 Transmission and Evolution and Ramifications for
  {COVID}-19 Diagnosis, Vaccine, and Medicine}.
\newblock {\em Journal of Chemical Information and Modeling}, 2020.
\newblock PMID: 32530284.

\bibitem{wang2020host}
Rui Wang, Yuta Hozumi, Yong-Hui Zheng, Changchuan Yin, and Guo-Wei Wei.
\newblock Host immune response driving {SARS-CoV-2} evolution.
\newblock {\em Viruses}, 12(10):1095, 2020.

\bibitem{wang2020topology}
Menglun Wang, Zixuan Cang, and Guo-Wei Wei.
\newblock A topology-based network tree for the prediction of protein--protein
  binding affinity changes following mutation.
\newblock {\em Nature Machine Intelligence}, 2(2):116--123, 2020.

\bibitem{carlsson2009topology}
Gunnar Carlsson.
\newblock Topology and data.
\newblock {\em Bulletin of the American Mathematical Society}, 46(2):255--308,
  2009.

\bibitem{edelsbrunner2000topological}
Herbert Edelsbrunner, David Letscher, and Afra Zomorodian.
\newblock Topological persistence and simplification.
\newblock In {\em Proceedings 41st annual symposium on foundations of computer
  science}, pages 454--463. IEEE, 2000.

\bibitem{xia2014persistent}
Kelin Xia and Guo-Wei Wei.
\newblock Persistent homology analysis of protein structure, flexibility, and
  folding.
\newblock {\em International journal for numerical methods in biomedical
  engineering}, 30(8):814--844, 2014.

\bibitem{kucukkal2015structural}
Tugba~G Kucukkal, Marharyta Petukh, Lin Li, and Emil Alexov.
\newblock Structural and physico-chemical effects of disease and non-disease
  {nsSNPs} on proteins.
\newblock {\em Current opinion in structural biology}, 32:18--24, 2015.

\bibitem{yue2005loss}
Peng Yue, Zhaolong Li, and John Moult.
\newblock Loss of protein structure stability as a major causative factor in
  monogenic disease.
\newblock {\em Journal of molecular biology}, 353(2):459--473, 2005.

\bibitem{sanjuan2016mechanisms}
Rafael Sanju{\'a}n and Pilar Domingo-Calap.
\newblock Mechanisms of viral mutation.
\newblock {\em Cellular and Molecular Life Sciences}, 73(23):4433--4448, 2016.

\bibitem{grubaugh2020making}
Nathan~D Grubaugh, William~P Hanage, and Angela~L Rasmussen.
\newblock Making sense of mutation: what {D614G} means for the {COVID-19}
  pandemic remains unclear.
\newblock {\em Cell}, 2020.

\bibitem{shu2017gisaid}
Yuelong Shu and John McCauley.
\newblock {GISAID}: {G}lobal initiative on sharing all influenza data--from
  vision to reality.
\newblock {\em Eurosurveillance}, 22(13), 2017.

\bibitem{chi2020neutralizing}
Xiangyang Chi, Renhong Yan, Jun Zhang, Guanying Zhang, Yuanyuan Zhang, Meng
  Hao, Zhe Zhang, Pengfei Fan, Yunzhu Dong, Yilong Yang, et~al.
\newblock A neutralizing human antibody binds to the {N}-terminal domain of the
  {S}pike protein of {SARS-CoV-2}.
\newblock {\em Science}, 369(6504):650--655, 2020.

\bibitem{wang2016raptorx}
Sheng Wang, Wei Li, Shiwang Liu, and Jinbo Xu.
\newblock Raptorx-property: a web server for protein structure property
  prediction.
\newblock {\em Nucleic acids research}, 44(W1):W430--W435, 2016.

\bibitem{wang2021structure}
Nan Wang, Yao Sun, Rui Feng, Yuxi Wang, Yan Guo, Li~Zhang, Yong-Qiang Deng, Lei
  Wang, Zhen Cui, Lei Cao, et~al.
\newblock Structure-based development of human antibody cocktails against
  sars-cov-2.
\newblock {\em Cell research}, 31(1):101--103, 2021.

\bibitem{acharya2020glycan}
Priyamvada Acharya, Wilton Williams, Rory Henderson, Katarzyna Janowska, Kartik
  Manne, Robert Parks, Margaret Deyton, Jordan Sprenz, Victoria Stalls, Megan
  Kopp, et~al.
\newblock A glycan cluster on the sars-cov-2 spike ectodomain is recognized by
  fab-dimerized glycan-reactive antibodies.
\newblock {\em bioRxiv}, 2020.

\bibitem{li2020potent}
Dianfan Li, Tingting Li, Hongmin Cai, Hebang Yao, Bingjie Zhou, Yapei Zhao,
  Wenming Qin, Cedric~AJ Hutter, Yanling Lai, Juan Bao, et~al.
\newblock Potent synthetic nanobodies against {SARS-CoV-2} and molecular basis
  for neutralization.
\newblock {\em bioRxiv}, 2020.

\bibitem{lan2020structure}
Jun Lan, Jiwan Ge, Jinfang Yu, Sisi Shan, Huan Zhou, Shilong Fan, Qi~Zhang,
  Xuanling Shi, Qisheng Wang, Linqi Zhang, et~al.
\newblock Structure of the {SARS-CoV-2} spike receptor-binding domain bound to
  the {ACE2} receptor.
\newblock {\em Nature}, pages 1--6, 2020.

\bibitem{yuan2020structural}
Meng Yuan, Hejun Liu, Nicholas~C Wu, Chang-Chun~D Lee, Xueyong Zhu, Fangzhu
  Zhao, Deli Huang, Wenli Yu, Yuanzi Hua, Henry Tien, et~al.
\newblock Structural basis of a shared antibody response to {SARS-CoV-2}.
\newblock {\em Science}, 369(6507):1119--1123, 2020.

\bibitem{wu2020noncompeting}
Yan Wu, Feiran Wang, Chenguang Shen, Weiyu Peng, Delin Li, Cheng Zhao, Zhaohui
  Li, Shihua Li, Yuhai Bi, Yang Yang, et~al.
\newblock A noncompeting pair of human neutralizing antibodies block {COVID-19}
  virus binding to its receptor {ACE2}.
\newblock {\em Science}, 2020.

\bibitem{piccoli2020mapping}
Luca Piccoli, Young-Jun Park, M~Alejandra Tortorici, Nadine Czudnochowski,
  Alexandra~C Walls, Martina Beltramello, Chiara Silacci-Fregni, Dora Pinto,
  Laura~E Rosen, John~E Bowen, et~al.
\newblock Mapping neutralizing and immunodominant sites on the sars-cov-2 spike
  receptor-binding domain by structure-guided high-resolution serology.
\newblock {\em Cell}, 183(4):1024--1042, 2020.

\bibitem{shi2020human}
Rui Shi, Chao Shan, Xiaomin Duan, Zhihai Chen, Peipei Liu, Jinwen Song, Tao
  Song, Xiaoshan Bi, Chao Han, Lianao Wu, et~al.
\newblock A human neutralizing antibody targets the receptor binding site of
  {SARS-CoV-2}.
\newblock {\em Nature}, pages 1--8, 2020.

\bibitem{rujas2020multivalency}
Edurne Rujas, Iga Kucharska, Yong~Zi Tan, Samir Benlekbir, Hong Cui, Tiantian
  Zhao, Gregory~A Wasney, Patrick Budylowski, Furkan Guvenc, Jocelyn~C Newton,
  et~al.
\newblock Multivalency transforms sars-cov-2 antibodies into broad and
  ultrapotent neutralizers.
\newblock {\em bioRxiv}, 2020.

\bibitem{huo2020structural}
Jiangdong Huo, Audrey Le~Bas, Reinis~R Ruza, Helen~ME Duyvesteyn, Halina
  Mikolajek, Tomas Malinauskas, Tiong~Kit Tan, Pramila Rijal, Maud Dumoux,
  Philip~N Ward, et~al.
\newblock Structural characterisation of a nanobody derived from a na{\"\i}ve
  library that neutralises sars-cov-2.
\newblock {\em Nature Protfolio}, 2020.

\bibitem{hurlburt2020structural}
Nicholas~K Hurlburt, Yu-Hsin Wan, Andrew~B Stuart, Junli Feng, Andrew~T
  McGuire, Leonidas Stamatatos, and Marie Pancera.
\newblock Structural basis for potent neutralization of {SARS-CoV-2} and role
  of antibody affinity maturation.
\newblock {\em bioRxiv}, 2020.

\bibitem{cao2020potent}
Yunlong Cao, Bin Su, Xianghua Guo, Wenjie Sun, Yongqiang Deng, Linlin Bao,
  Qinyu Zhu, Xu~Zhang, Yinghui Zheng, Chenyang Geng, et~al.
\newblock Potent neutralizing antibodies against {SARS-CoV-2} identified by
  high-throughput single-cell sequencing of convalescent patients’ {B} cells.
\newblock {\em Cell}, 2020.

\bibitem{pinto2020structural}
Dora Pinto, Young-Jun Park, Martina Beltramello, Alexandra~C Walls, M~Alejandra
  Tortorici, Siro Bianchi, Stefano Jaconi, Katja Culap, Fabrizia Zatta, Anna
  De~Marco, et~al.
\newblock Structural and functional analysis of a potent sarbecovirus
  neutralizing antibody.
\newblock {\em BioRxiv}, 2020.

\bibitem{zhou2020structural}
Daming Zhou, Helen~ME Duyvesteyn, Cheng-Pin Chen, Chung-Guei Huang, Ting-Hua
  Chen, Shin-Ru Shih, Yi-Chun Lin, Chien-Yu Cheng, Shu-Hsing Cheng, Yhu-Chering
  Huang, et~al.
\newblock Structural basis for the neutralization of {SARS-CoV-2} by an
  antibody from a convalescent patient.
\newblock {\em Nature Structural \& Molecular Biology}, pages 1--9, 2020.

\bibitem{du2020structurally}
Shuo Du, Yunlong Cao, Qinyu Zhu, Pin Yu, Feifei Qi, Guopeng Wang, Xiaoxia Du,
  Linlin Bao, Wei Deng, Hua Zhu, et~al.
\newblock Structurally resolved sars-cov-2 antibody shows high efficacy in
  severely infected hamsters and provides a potent cocktail pairing strategy.
\newblock {\em Cell}, 183(4):1013--1023, 2020.

\bibitem{lv2020structural}
Zhe Lv, Yong-Qiang Deng, Qing Ye, Lei Cao, Chun-Yun Sun, Changfa Fan, Weijin
  Huang, Shihui Sun, Yao Sun, Ling Zhu, et~al.
\newblock Structural basis for neutralization of {SARS-CoV-2} and {SARS-CoV} by
  a potent therapeutic antibody.
\newblock {\em Science}, 369(6510):1505--1509, 2020.

\bibitem{wu2020alternative}
Nicholas~C Wu, Meng Yuan, Hejun Liu, Chang-Chun~D Lee, Xueyong Zhu, Sandhya
  Bangaru, Jonathan~L Torres, Tom~G Caniels, Philip~JM Brouwer, Marit~J
  Van~Gils, et~al.
\newblock An alternative binding mode of {IGHV3-53} antibodies to the
  {SARS-CoV-2} receptor binding domain.
\newblock {\em BioRxiv}, 2020.

\bibitem{ju2020human}
Bin Ju, Qi~Zhang, Jiwan Ge, Ruoke Wang, Jing Sun, Xiangyang Ge, Jiazhen Yu,
  Sisi Shan, Bing Zhou, Shuo Song, et~al.
\newblock Human neutralizing antibodies elicited by {SARS-CoV-2} infection.
\newblock {\em Nature}, pages 1--8, 2020.

\bibitem{kreye2020therapeutic}
Jakob Kreye, S~Momsen Reincke, Hans-Christian Kornau, Elisa S{\'a}nchez-Sendin,
  Victor~Max Corman, Hejun Liu, Meng Yuan, Nicholas~C Wu, Xueyong Zhu,
  Chang-Chun~D Lee, et~al.
\newblock A therapeutic non-self-reactive sars-cov-2 antibody protects from
  lung pathology in a covid-19 hamster model.
\newblock {\em Cell}, 183(4):1058--1069, 2020.

\bibitem{tortorici2020ultrapotent}
M~Alejandra Tortorici, Martina Beltramello, Florian~A Lempp, Dora Pinto, Ha~V
  Dang, Laura~E Rosen, Matthew McCallum, John Bowen, Andrea Minola, Stefano
  Jaconi, et~al.
\newblock Ultrapotent human antibodies protect against sars-cov-2 challenge via
  multiple mechanisms.
\newblock {\em Science}, 370(6519):950--957, 2020.

\bibitem{bracken2021bi}
Colton~J Bracken, Shion~A Lim, Paige Solomon, Nicholas~J Rettko, Duy~P Nguyen,
  Beth~Shoshana Zha, Kaitlin Schaefer, James~R Byrnes, Jie Zhou, Irene Lui,
  et~al.
\newblock Bi-paratopic and multivalent vh domains block ace2 binding and
  neutralize sars-cov-2.
\newblock {\em Nature Chemical Biology}, 17(1):113--121, 2021.

\bibitem{liu2020cross}
Hejun Liu, Nicholas~C Wu, Meng Yuan, Sandhya Bangaru, Jonathan~L Torres, Tom~G
  Caniels, Jelle Van~Schooten, Xueyong Zhu, Chang-Chun~D Lee, Philip~JM
  Brouwer, et~al.
\newblock Cross-neutralization of a sars-cov-2 antibody to a functionally
  conserved site is mediated by avidity.
\newblock {\em Immunity}, 53(6):1272--1280, 2020.

\bibitem{clark2020molecular}
Sarah~Ashley Clark, Lars~Eric Clark, Junhua Pan, Adrian Coscia, Lindsay~GA
  McKay, Sundaresh Shankar, Rebecca~I Johnson, Anthony Griffiths, and Jonathan
  Abraham.
\newblock Molecular basis for a germline-biased neutralizing antibody response
  to sars-cov-2.
\newblock {\em bioRxiv}, 2020.

\bibitem{bertoglio2020sars}
Federico Bertoglio, Viola F{\"u}hner, Maximilian Ruschig, Philip~Alexander
  Heine, Ulfert Rand, Thomas Kl{\"u}nemann, Doris Meier, Nora Langreder,
  Stephan Steinke, Rico Ballmann, et~al.
\newblock A sars-cov-2 neutralizing antibody selected from covid-19 patients by
  phage display is binding to the ace2-rbd interface and is tolerant to known
  rbd mutations.
\newblock {\em bioRxiv}, 2020.

\bibitem{custodio2020selection}
T{\^a}nia~F Cust{\'o}dio, Hrishikesh Das, Daniel~J Sheward, Leo Hanke, Samuel
  Pazicky, Joanna Pieprzyk, Mich{\`e}le Sorgenfrei, Martin~A Schroer, Andrey~Yu
  Gruzinov, Cy~M Jeffries, et~al.
\newblock Selection, biophysical and structural analysis of synthetic
  nanobodies that effectively neutralize sars-cov-2.
\newblock {\em Nature communications}, 11(1):1--11, 2020.

\bibitem{yao2021rational}
Hangping Yao, Yao Sun, Yong-Qiang Deng, Nan Wang, Yongcong Tan, Na-Na Zhang,
  Xiao-Feng Li, Chao Kong, Yan-Peng Xu, Qi~Chen, et~al.
\newblock Rational development of a human antibody cocktail that deploys
  multiple functions to confer pan-sars-covs protection.
\newblock {\em Cell research}, 31(1):25--36, 2021.

\bibitem{tian2020potent}
Xiaolong Tian, Cheng Li, Ailing Huang, Shuai Xia, Sicong Lu, Zhengli Shi,
  Lu~Lu, Shibo Jiang, Zhenlin Yang, Yanling Wu, et~al.
\newblock Potent binding of 2019 novel coronavirus spike protein by a {SARS}
  coronavirus-specific human monoclonal antibody.
\newblock {\em Emerging microbes \& infections}, 9(1):382--385, 2020.

\bibitem{ter2006human}
Jan Ter~Meulen, Edward~N Van Den~Brink, Leo~LM Poon, Wilfred~E Marissen,
  Cynthia~SW Leung, Freek Cox, Chung~Y Cheung, Arjen~Q Bakker, Johannes~A
  Bogaards, Els Van~Deventer, et~al.
\newblock Human monoclonal antibody combination against {SARS} coronavirus:
  synergy and coverage of escape mutants.
\newblock {\em PLoS medicine}, 3(7), 2006.

\bibitem{huang2010cd}
Ying Huang, Beifang Niu, Ying Gao, Limin Fu, and Weizhong Li.
\newblock Cd-hit suite: a web server for clustering and comparing biological
  sequences.
\newblock {\em Bioinformatics}, 26(5):680--682, 2010.

\bibitem{liu2020potent}
Lihong Liu, Pengfei Wang, Manoj~S Nair, Jian Yu, Micah Rapp, Qian Wang, Yang
  Luo, Jasper F-W Chan, Vincent Sahi, Amir Figueroa, et~al.
\newblock Potent neutralizing antibodies against multiple epitopes on
  {SARS-CoV-2} spike.
\newblock {\em Nature}, 584(7821):450--456, 2020.

\bibitem{chen2020mutations}
Jiahui Chen, Rui Wang, Menglun Wang, and Guo-Wei Wei.
\newblock Mutations strengthened {SARS-CoV-2} infectivity.
\newblock {\em Journal of Molecular Biology}, 2020.

\bibitem{chen2020sars}
Wen-Hsiang Chen, Ulrich Strych, Peter~J Hotez, and Maria~Elena Bottazzi.
\newblock The {SARS-CoV-2} vaccine pipeline: an overview.
\newblock {\em Current tropical medicine reports}, pages 1--4, 2020.

\bibitem{lin2007safety}
J~Lin, Jian-San Zhang, Nan Su, Jian-Guo Xu, Nan Wang, Jiang-Ting Chen, Xin
  Chen, Yu-Xuan Liu, Hong Gao, Yu-Ping Jia, et~al.
\newblock Safety and immunogenicity from a phase {I} trial of inactivated
  severe acute respiratory syndrome coronavirus vaccine.
\newblock {\em Antiviral therapy}, 12(7):1107, 2007.

\bibitem{li2013bioinformatic}
Yanhua Li, Xianfei Liu, Yuejie Zhu, Xiaotao Zhou, Chunbao Cao, Xiaoan Hu,
  Haimei Ma, Hao Wen, Xiumin Ma, and Jian-Bing Ding.
\newblock Bioinformatic prediction of epitopes in the {Emy162} antigen of
  {E}chinococcus multilocularis.
\newblock {\em Experimental and therapeutic medicine}, 6(2):335--340, 2013.

\bibitem{kringelum2013structural}
Jens~Vindahl Kringelum, Morten Nielsen, S{\o}ren~Berg Padkj{\ae}r, and Ole
  Lund.
\newblock Structural analysis of {B}-cell epitopes in antibody: protein
  complexes.
\newblock {\em Molecular immunology}, 53(1-2):24--34, 2013.

\bibitem{linsky2020novo}
Thomas~W Linsky, Renan Vergara, Nuria Codina, Jorgen~W Nelson, Matthew~J
  Walker, Wen Su, Christopher~O Barnes, Tien-Ying Hsiang, Katharina
  Esser-Nobis, Kevin Yu, et~al.
\newblock De novo design of potent and resilient hace2 decoys to neutralize
  sars-cov-2.
\newblock {\em Science}, 370(6521):1208--1214, 2020.

\end{thebibliography}

\end{document}